\documentclass[12pt,onecolumn,oneside]{aa}
\usepackage{graphicx}
\usepackage{psfig}
\usepackage{epsf}
\usepackage{txfonts}
\def\farcs{\hbox{$.\!\!^{\prime\prime}$}}
\def\fsecs{\hbox{$.\!\!^{\rm s}$}}

\newcounter{saveeqn}

\newcommand{\alpheqn}{\addtocounter{equation}{1}\setcounter{saveeqn}{\value{equation}}%
\setcounter{equation}{0}%
\renewcommand{\theequation}{\arabic{saveeqn} \alph{equation}}}
\newcommand{\reseteqn}{\setcounter{equation}{\value{saveeqn}}%
\renewcommand{\theequation}{\arabic{equation}}}

\newcommand{\mylabel}[2]{\begin{eqnarray}\label{#1} \nonumber \end{eqnarray} \vspace*{#2}}

\begin{document}

\thispagestyle{empty}


\begin{center} 
   {\Huge \bf The G9.62+0.19--F Hot Molecular Core} \bigskip\\

   {\Large \bf The Infrared View on Very Young Massive Stars} \bigskip\\
   
   \vspace{1cm}

   {\large \bf Hendrik Linz (1,2), Bringfried Stecklum (1),
          Thomas Henning (2), \smallskip\\
	  Peter Hofner (3,4), and
          Bernhard Brandl (5,6)
          } \bigskip \\

    \vspace{2cm}


         { \large \bf 
	    (1)  Th\"uringer Landessternwarte Tautenburg, 
                 Sternwarte 5, D--07778 Tautenburg, Germany \\
                 email: linz@tls--tautenburg.de, stecklum@tls--tautenburg.de \smallskip\\
            (2) Max--Planck--Institut f\"ur Astronomie, 
	        K\"onigstuhl 17, D--69117 Heidelberg, Germany \\
	        email: henning@mpia-hd.mpg.de  \smallskip\\
            (3) Physics Department, New Mexico Tech, 801 Leroy Place, Socorro, N.M. 87801, USA \\
	    (4) NRAO, PO Box 0, Socorro, N.M. 87801, USA \\
	        email: phofner@nrao.edu  \smallskip\\
            (5) Center for Radiophysics \& Space Research, Cornell University, Ithaca, NY 14853, USA   \smallskip \\     
            (6) Sterrewacht Leiden, Niels Bohr Weg 2 (\#535), PO Box 9513, 2300 RA Leiden, Netherlands \\
                email: brandl@strw.leidenuniv.nl         
             }
\end{center}

\vspace{2cm}

accepted for publication in {\sl Astronomy \& Astrophysics}



\newpage

\begin{center}
   {\LARGE \bf The G9.62+0.19--F Hot Molecular Core} \bigskip\\

   {\Large \bf The Infrared View on Very Young Massive Stars \footnote{Based 
   on observations made with the ESO VLT at the Paranal Observatory under 
   programme IDs 63.I-0329 and 67.C-0264
    and with ESO'S TIMMI2 on La Silla 
   under programme ID 71.C-0438.}} \bigskip\\
   
   {\large  H. Linz, B. Stecklum, Th. Henning, P. Hofner, \& B. Brandl} \bigskip\\

\vspace{1cm}
   
   {\large \sf Abstract} \bigskip\\

   \parbox{14cm}{ \sf We present the results of an extensive infrared study of the massive star--forming region G9.62+0.19.
   The data cover information from broad-- and narrow--band filters in the wavelength range from 1 to 19 $\mu$m 
   and are obtained with ESO's near--
   and thermal infrared camera ISAAC at the VLT and  with the mid--infrared cameras TIMMI2 (La Silla, ESO) and
   SpectroCam-10 (Mt. Palomar). The high sensitivity and resolution provided by these facilities revealed 
   intriguing new details
   of this star--forming region and especially about the embedded hot molecular core (HMC) -- component F. We 
   analyse the newly found infrared sub--structure  of four objects in this HMC region. 
   While one of these objects (F2) is
   probably a foreground field star, the nature of the brightest object in the near--infrared there (F1) 
   remains somewhat enigmatic. Our 
   new astrometry proves that this object is not coincident with the peak of the molecular line emission of
   the HMC, but displaced by $\sim$ 1.7 arcsecs which translates to nearly 10000 AU on a linear scale. 
   On the basis of the available data we estimate this object to be an additional embedded object with a 
   dense dust shell. Very near the HMC location we find  L' band emission which strongly 
   rises in flux towards longer wavelengths. We presume that this emission (F4)
   arises from the envelope of the HMC which is known to be associated with  a molecular 
   outflow roughly aligned along the line of sight. Thus, the clearing effect of this outflow causes 
   strong deviations from spherical symmetry which might allow 
   infrared emission from the HMC to escape through the outflow cavities. 
   This presents the first direct detection of an HMC at a wavelength as short as 3.8 $\mu$m. 
    At 11.7 $\mu$m and 18.75 $\mu$m, the HMC counterpart F4 ultimately proves to be the most
   luminous IR source within the G9.62+0.19--F region.\\
   In addition, within the entire G9.62+0.19 
   complex our narrow--band data and the K band imaging polarimetry 
   reveal well--defined regions
   of enhanced Br$\gamma$ and H$_2$ emission as well as a sector where a large contribution comes from
   scattered light. In combination with high--resolution radio data we make predictions about the
   extinction within this star--forming region which clarifies why some of the associated ultracompact
   H{\sc ii} regions are not visible in the near--infrared.\\
   Our investigations show the complexity of massive star formation in full grandeur, but they also
   demonstrate that the related problems can be tackled by observations using the new generation of 
   infrared cameras.
   }
  
  
\end{center}


\section{Introduction}

   During the last 10 to 15 years, a canonical pictu\-re of the formation
   of low--mass stars has been established (Shu et al. \cite{Shu}, Andr\'e et al. 
   \cite{Andre}),  although details of the formation process are still under discussion (e.g.,
   Hartmann \cite{Hartmann}). 
   On the other hand, a corresponding consistent model for the birth of massive
   stars (M $\ge$ 8--10 M$_{\odot}$) has still to be developed
   (see, e.g, reviews by Kurtz et al. \cite{Kurtz1}, Stahler et al. \cite{Stahler}, Yorke \cite{Yorke2}).
   On average, sites of high--mass star formation are found at larger distances
   from the Sun than regions of low--mass star formation. In addition, they are located 
   in more crowded and more heavily obscured regions than young low--mass
   stars.  Both facts strongly hamper observational investigations. Furthermore, the evolution
   from a contracting cloud core to a high--mass main sequence star occurs on 
   much shorter time scales, whereby the protostar shows more intensive interaction 
   with the natal environment  which further increases the complexity of the problem. Thus, the 
   requirements to study massive star formation are severe  for both, the observer
   and the theoretician (Garay \& Lizano \cite{Garay2}, Henning et al. \cite{Henning}).\\
   Over the last decade, it has become possible to explore earlier stages in the evolution
   of massive young stellar objects (YSOs). While for many years ultracompact H{\small II} 
   regions (UCH{\small II}s) played the 
   central role in these investigations (e.g., Wood \& Churchwell \cite{Wood}, Kurtz et al.
   \cite{Kurtz0}), recent studies have revealed
   that warm (T $\ge$ 100 K), dense (n$_{\rm H_2}$ = 10$^6 . . 10^8 \,
   $cm$^{-3}$) and compact (size $\le$ 0.1 pc) condensations within the molecular 
   clouds, the so--called hot molecular cores (HMCs), can provide insights into even earlier 
   phases of the formation of high--mass stars (Cesaroni et al. \cite{Cesaroni1}, Kurtz et al. 
   \cite{Kurtz1}). 
   
   It was found that HMCs are close to, but  often not coincident with, the adjacent 
   UCH{\small II}s (Cesaroni et al. \cite{Cesaroni1}). 
   Several authors have proposed an evolutionary scenario leading from HMCs to
   UCH{\small II}s and, finally, to more evolved H{\small II} regions
   (Cesaroni et al. \cite{Cesaroni1}, Kurtz et al. \cite{Kurtz1}). 
   
   The well--studied galactic region G9.62+0.19
   (Garay et al. \cite{Garay1}, Hofner et al. \cite{Hofner2}) at a distance of $\sim$ 5.7 kpc 
   (Hofner et al. \cite{Hofner1}) contains a  
   number of massive YSOs in different evolutionary stages located close to each 
   other. In this respect, there are indications that an age gradient goes from western 
   (older) regions to eastern (younger) ones (Hofner et al. \cite{Hofner1}, \cite{Hofner2}, 
   Testi et al. \cite{Testi1}). One of the 
   most interesting objects of the region (source F) is confirmed to be an HMC 
   (Cesaroni et al. \cite{Cesaroni1}, Hofner et al. \cite{Hofner1}, \cite{Hofner2}) by means of 
   interferometric observations
   of the dense gas tracers NH$_3$ and CH$_3$CN. Situated between two 
   UCH{\small II}s (radio components D and E), weak cm--continuum radiation 
   was found recently for component F using high--sensitivity {\sl VLA} 
   measurements (Testi et al. \cite{Testi2}).
   
   Another feature, which makes this HMC a unique target, is the fact that very close
   to the location of G9.62+0.19--F emission in the 2.2 $\mu$m continuum (K band) was detected 
   (Testi et al. \cite{Testi1}).
   Corresponding to the standard models of HMCs, describing them as 
   spherically symmetric and extremely dense dust and gas spheres (Kaufman et al. \cite{Kaufman}, 
   Osorio et al. \cite{Osorio}),
   the optical depth should be so high that the HMC should remain undetectable at near-- and
   mid--infrared wavelengths. This raises the question, what the actual origin of the K band emission is.
   In this respect it is important to mention that Hofner et al. (\cite{Hofner4}) have detected  a 
   molecular outflow 
   in G9.62+0.19 with the flow orientation very close to the line of sight,
   for which they favour component F as the most likely driving source. The working hypothesis  of
   Hofner et al. (\cite{Hofner4}) was
   that the outflow forms a cavity of strongly decreased density roughly along the line
   of sight through which the infrared (IR) emission from the HMC can escape.
   
   While the G9.62+0.19 region was covered over the years by many radio observations with more and
   more increased resolution and sensitivity, detailed information about this star--forming region 
   in the infrared is rather scarce.
   By utilising high--resolution near-- and mid--infrared data, we have explored 
   the IR properties of this region. In Sect.~2 we explain details of the observations
   we have conducted during the last  four years. In Sect.~3 we present the direct results of these
   measurements and discuss the global features of the region and the specifics of 
   the HMC zone on which the emphasis of our analysis will lie. Since 
   an accurate astrometry will be a crucial issue for
   the following interpretation of the data, we included a detailed description of the
   astrometric techniques we have applied. In combination with 
   interferometric radio maps and 
   a radiative 
   transfer model, we discuss the implications of our infrared data for an interpretation
   of the objects in the HMC region in Sect.~4. Furthermore, we compare the HMC region to the
   UCH{\small II}s nearby and to the HMC in W3(H$_2$O).
   Finally, in Sect.~5 we summarise our conclusions.


\section{Infrared observations}

   \begin{table*}[htb]
      \caption[]{Summary of observations}
         \label{obstable}

         \begin{tabular}{lrcrcc}        
            \hline
	    \hline
            \noalign{\smallskip}
  Date      &  Filter \ and \ $\lambda_{\rm c}$ & Telescope / Camera   &  FOV$^{\, \rm a}$ &  PSF FWHM  &  Ref. Star\\
            \noalign{\smallskip}
            \hline
            \noalign{\smallskip}
  1999 Sep  & J     \ \ \	  1.25  $\mu$m  &  ANTU / ISAAC        &  151$''$   &  0\farcs 55  &  S234--E, S071--D \\
  1999 Sep  & H     \ \ \	  1.65  $\mu$m  &      "	       &  151$''$   &  0\farcs 65	&  S234--E, S071--D \\
  1999 Sep  & Ks    \ \ \         2.16  $\mu$m  &      "	       &  151$''$   &  0\farcs 55	&  S234--E, S071--D \\
  2001 May  & nb\_K \ \ \	  2.09  $\mu$m  &      "	       &  $^{\rm b}\, 151''$ 	   &  0\farcs 65  &   \\
  2001 May  & H$_2$(1--0)S1 \ \   2.122 $\mu$m  &      "	       &  151$''$  &  0\farcs 48  &   \\ 
  2001 May  & Br$\, \gamma$ \ \   2.166 $\mu$m  &      "	       &  151$''$  &  0\farcs 44  &   \\
  2001 Jun  & L$'$	    \ \ \ 3.78  $\mu$m  &      "	       &   79$''$  &  0\farcs 55 $\times$ 0\farcs 42  &  HR 6070 \\
  2001 Jun  & Br$\, \alpha$ \ \ \ 4.07  $\mu$m  &      "	       &   79$''$  &  0\farcs 48 $\times$ 0\farcs 41  &  HR 6070 \\
  2001 Jun  & nb\_M	    \ \ \ 4.66  $\mu$m  &      "	       &   79$''$  &  0\farcs 47 $\times$ 0\farcs 40  &  HR 6070 \\
  2003 Jul  & N1          \ \ \   8.70  $\mu$m  &  ESO 3.6-m / TIMMI2  &  $64'' \times 48''$  & 0\farcs 86 &  HD 169916	 \\	       
  1999 Jun  & N 	  \ \ \  11.70  $\mu$m  &  5-m Hale / SC-10    &   16$''$  &  0\farcs 76   & $\delta$ Oph  \\
  2003 Jul  & Q2          \ \ \  18.75  $\mu$m  &  ESO 3.6-m / TIMMI2  &  $64'' \times 48''$  & 1\farcs 60  &  HD 187642 \\	       

            \noalign{\smallskip}
            \hline
         \end{tabular} 
\begin{list}{}{}
\item $^{\, \rm a}$ \ Field of view for single frames, respectively 
\item $^{\, \rm b}$ \ Imaging polarimetry with Wollaston prism and slit mask in the optical path
\end{list}
   \end{table*}

\subsection{J, H, and Ks band observations}

   The near--infrared data were obtained in September 1999 with the infrared camera {\sl ISAAC} 
   (Moorwood \cite{Moorwood}) at the
   {\sl VLT ANTU} telescope 
   within the programme 
   ID 63.I-0329. We used the opportunity to utilise these archived data for our purposes. \\ 
   The following NIR broad--band filters were used:
   J ($\lambda _{c} = 1.25\,\mu$m, FWHM = 0.29 $\mu$m),
   H ($\lambda _{c} = 1.65\, \mu$m, FWHM = 0.30 $\mu$m),
   K$_{\mathrm s}$ ($\lambda _{c} = 2.16\, \mu$m, FWHM = 0.27 $\mu$m).
   The seeing conditions were between 0.65 arcsec (H--band) and 0.55 arcsec (J-- and 
   K$_{\mathrm s}$--band), respectively. The pixel scale is 0$.\!\!''147$/px 
   which results in a field of approximately $2.\!\!'5 \times 2.\!\!'5$. Within the overlap area 
   of the 5 offset images per wavelength, the total integration time 
   on source for each filter was 266 s . After the standard procedure (dark--field and
   flat--field reduction, bad pixel removal and sky subtraction), utilising our own 
   {\sl IDL}\footnote{Interactive Data Language}--based
   reduction pipeline for the {\sl ISAAC} data (see also  Stecklum et al. \cite{Stecklum4} 
   and Apai et al. \cite{Apai}), we corrected for
   the field distortion by applying the correction terms recently provided by ESO. 
   Then the 5 dithered frames were combined to a larger mosaic image achieving
   subpixel accuracy. As standards for photometric calibration, the stars S234--E and
   S071--D\footnote{ We mention here that, unlike the other finding charts shown in 
   Persson et al. (\cite{Persson}), the finding chart for S071--D is (accidentally)
   not centred on the actual calibration star which at first caused some confusion for our calibration.} 
   (from the list of faint NIR standard stars by Persson et al. \cite{Persson}) were 
   observed before and after the observation of G9.62+0.19, respectively. 
   
\subsection{Br $\gamma$ and H$_2$ observation 
   and K narrow--band polarimetry
   }\label{pol_obs}

   Images of G9.62+0.19 in the Br $\gamma$ ($\lambda _{c} = 2.16\, \mu$m, FWHM = 0.02 $\mu$m)
   and in the H$_2$(1--)S1 ($\lambda _{c} = 2.12\, \mu$m, FWHM = 0.02 $\mu$m) narrow--band
   filters were obtained in May, 2001, again with {\sl ISAAC} at the {\sl VLT ANTU} telescope.
   Pixel scale and field of view are the same as for the JHK$_{\mathrm s}$ broadband imaging.
   The total integration times on source for both filters were 500 s within the overlap area 
   of the 5 offset images per wavelength.
   
   The imaging polarimetry system of {\sl ISAAC} was used to obtain complementary NIR data 
   for the G9.62+0.19 region. The camera is equipped with a Wollaston prism which divides the 
   infalling light into two perpendicular polarised beams having a separation of $\sim \, 22 ''$.
   The prism is used with a special slot mask (3 opaque stripes, each being 20$''$ wide) to keep the
   two fully polarised copies of the imaged area separated on the detector. The first step is
   to simultaneously obtain the 0$^{\circ}$ and 90$^{\circ}$ data. Then, the reference plane
   is rotated by 45$^{\circ}$ by offsetting the image derotator of the alt--az mounting control system
   of the telescope accordingly. In this way, the 45$^{\circ}$ and 135$^{\circ}$ data can be obtained.   
   To ensure a seamless
   coverage of the entire star--forming region, we used a 3--point dither pattern 
   perpendicular to the stripe orientation. A narrow--band K filter was used ($\lambda _{c} = 2.09\, 
   \mu$m, FWHM = 0.02 $\mu$m), which does not contain strong spectral lines.

\subsection{L' and M band observations}

   {\sl ISAAC} is also capable of imaging in the L' ($\lambda _{c} = 3.78\,\mu$m, FWHM = 0.58 $\mu$m) and
    nb\_M ($\lambda _{c} = 4.66\,\mu$m, FWHM = 0.10 $\mu$m) thermal infrared narrow bands. Within our programme 
   67.C-0264, service time observations were performed for G9.62+0.19 in June, 2001, thus, 
   very shortly after the repair of the {\sl ISAAC} 
   chopping secondary mirror in April and May, 2001. The fact that during the first weeks after recommission 
   of the chopping mode the data were taken without field stabilisation resulted in a slightly
   degraded image quality, recognizable by means of slightly elongated star shapes.
   In addition to the above--mentioned filters the 4.07 $\mu$m narrow--band filter (FWHM = 0.07 $\mu$m) was
   used which includes the Brackett $\alpha$ line. The pixel scale in connection with 
   the {\sl ALADDIN} detector array is 0$.\!\!''071$/px, the field of view is then
   $73'' \times 73''$. Since the L' and nb\_M bands lie within the thermal infrared, on--source/off--source 
   chopping is required to remove the thermal background in the images. Therefore, we
   have the positive and the negative beam in the images, separated by 15 arcsec. 
   In addition, a 5 point dither pattern was applied. 
   The standard star HR 6070 (van der Bliek et al. \cite{Bliek}) was used to calibrate the data.

\subsection{N and Q band observations}

   We utilised two different cameras to image the central region of G9.62+0.19 in the
   mid--infrared (MIR).
   The 11.7 $\mu$m image was obtained in June 1999 using {\sl SpectroCam-10} (Hayward et al. \cite{Hayward}) 
   at the 200--inch Hale Telescope of the Palomar Observatory\footnote{Observations at 
   the Palomar Observatory were made as part of a continuing collaborative agreement 
   between the California Institute of Technology, Cornell University, and the Jet 
   Propulsion Laboratory.}.  The camera uses a 128$^2$ pixel Si:As BIBIB detector manufactored by
   Rockwell. For imaging the chip is binned into a $64 \times 64$ pixel array with 0\farcs 256/pixel.
   The effective wavelength of the filter was 11.7 $\mu$m with a FWHM of 1 $\mu$m.  Chopping and nodding was 
   performed in the standard way (chopping parallel to nodding) with a throw of 20 arcsec. After applying
   a standard chop--and--nod reduction for the raw frames,
   a 5--frame mosaic was combined into the final image. The average
   on--source integration time at each pixel is 200 s. The star $\delta$ Oph 
   (HD 146051) served as standard for the flux calibration. \\
    Second, we used the {\sl TIMMI2} camera (Reimann et al. \cite{Reimann}) at the ESO 3.6-m
   telescope for additional  MIR observations in July 2003. Equipped with  a 320 $\times$ 240
   px$^2$ Si:As BIB {\sl Raytheon} array, the FOV in the high-resolution mode is $64'' \times 48''$.
   Hereby the N1 filter ($\lambda_{\rm c} = 8.7 \,\mu$m, FWHM = 1 $\mu$m) and 
   the Q2 filter ($\lambda_{\rm c} = 18.75 \,\mu$m, FWHM = 0.9 $\mu$m) have been used. The chopping and
   nodding throws were 15 arcsec, but both movements are performed perpendicular when using the
   TIMMI2 standard setup. After combination of the chopped and nodded beams the total on--source times 
   are 749 s for N1 and 795 s for Q2, respectively. While the 
   sky was stable and almost photometric during the N1 measurements, the Q2 band observations on the
   following day suffered from mediocre atmospheric conditions. The stars HD 169916 (N1) and HD 187642 (Q2)
   from the list of TIMMI2 standard stars were used for calibration. We mention that the Q2 standard
   star was observed immediately after the end of the Q2 exposure for the science target und hence was 
   affected by the same sky conditions.
   
   With the {\sl TIMMI2} data we also tried out a different (and perhaps more sophisticated) approach to 
   create the final image, namely the so--called projected Landweber matrix restoration 
   (Bertero et al. \cite{Bertero} and references therein).
   We refer to appendix \ref{Landweber} for some comments about the restoration method itself and 
   the modifications we applied.\\
   Because of our modified chop/nod observational approach we have to deal with artifacts (see appendix)
   in two orthogonal directions. We have undertaken the first steps to reduce these collateral effects (e.g., 
   ghost images) by precisely aligning the array orientation along the chopping/nodding throws and by 
   carefully registering the real values of the throws. A simple way to minimize the multiple artifacts
   would be, according to Bertero et al. (\cite{Bertero}), to combine several restored images each having 
   different throws.  Due to time constraints, however, we could only perform one long integration for 
   each filter, so we were not able to completely 
   avoid these perturbing effects. In this respect one has to remember that a part of the original signal
   is then still contained in the artefacts. The effects on photometric accuracy are under investigation.
   For this paper we will use the restored {\sl TIMMI2} images only to reveal the MIR morphology 
   of the region. The photometry was always performed on unrestored and undeconvolved images.


\section{Results}

 \subsection{NIR data}\label{NIR data}

\begin{figure*}
   \centering
     \includegraphics[width=\textwidth,angle=0]{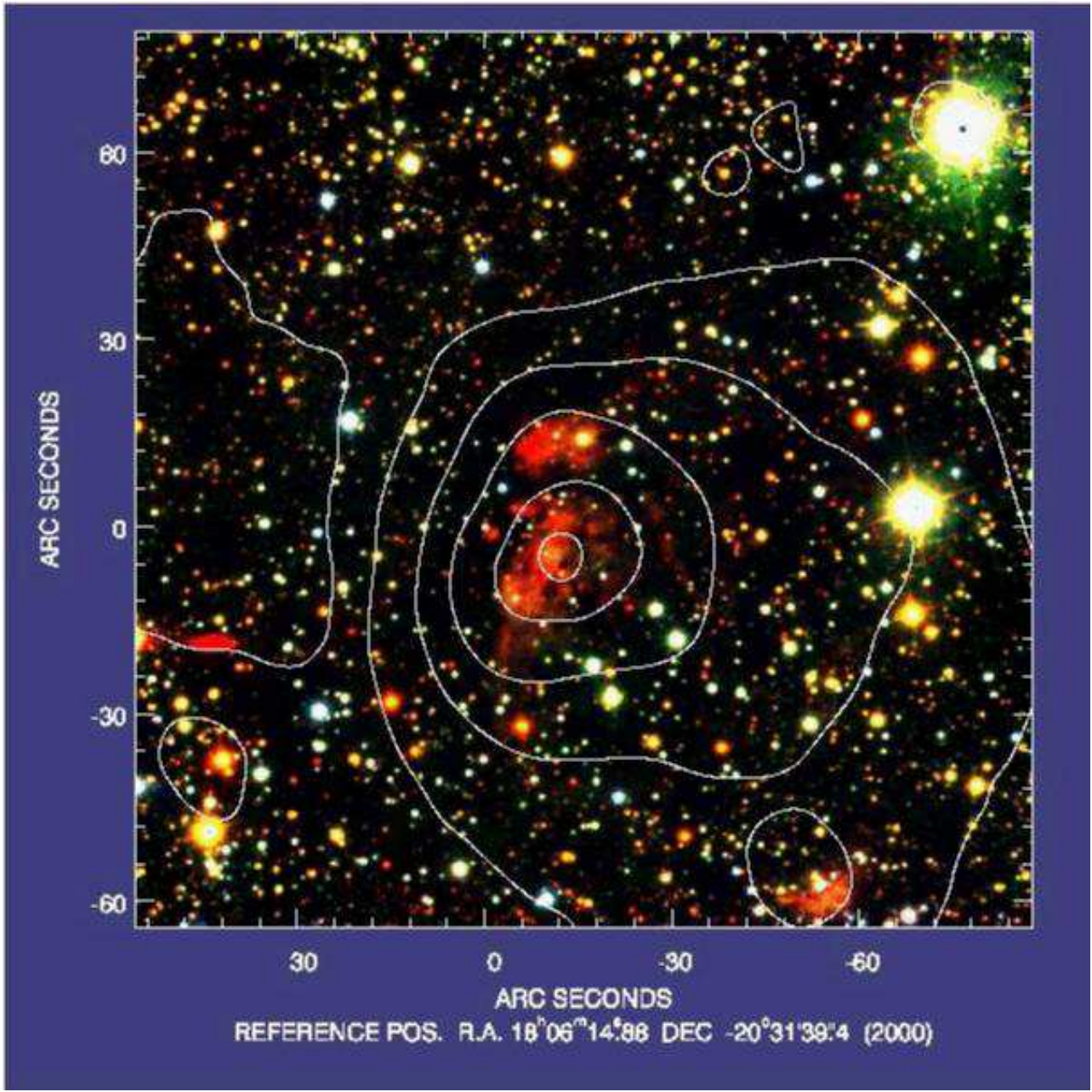} 
      \caption{Colour--coded image of the entire G9.62+0.19 region taken in the three 
  broad--band NIR filters J (blue), H (green), and  K$_{\mathrm s}$ (red). The large--scale 
  contour lines denote the emission levels derived from the 8.28 $\mu$m image 
  of the related {\sl MSX} source. The left--most large contour line indicates the position 
  of the close--by Infrared Dark Cloud.
              }
         \label{VLT1}
   \end{figure*}

\begin{figure*}
   \centering
     \includegraphics[width=\textwidth]{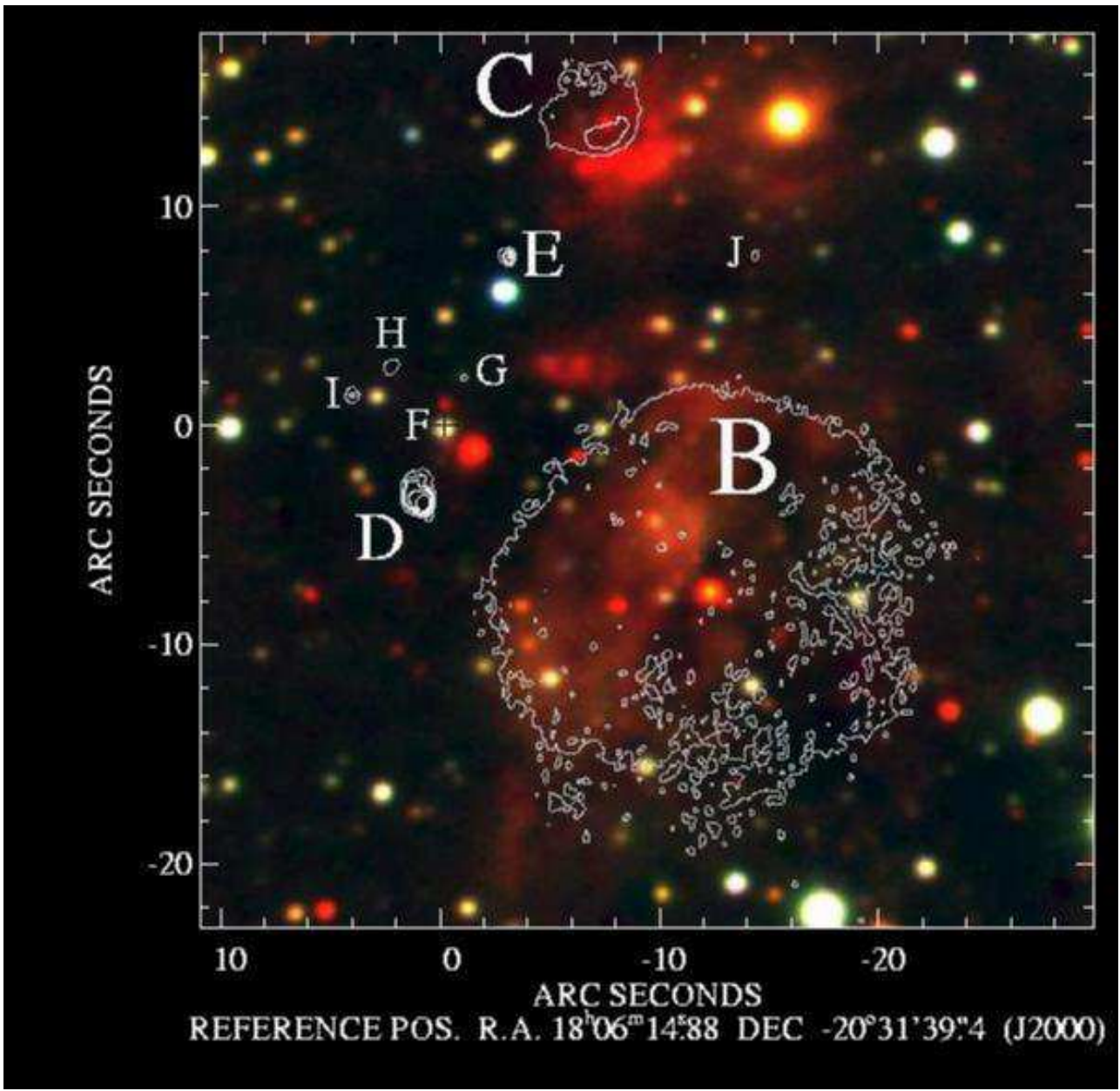}
      \caption{Cutout of the central region of Fig. \ref{VLT1}. The overlaid contour lines denote the
               3.6 cm emission of the ionized regions as measured by Testi et al. (\cite{Testi2}). The source 
               annotation follows their nomenclature. Note, that also component F features weak
	       3.6 cm emission -- the related contour line merges with the seeing disk of the
	       yellow star. The black cross marks the peak of the NH$_3$(5,5)
	       HMC emission (Hofner et al. \cite{Hofner1}).
              }
         \label{VLT2}
   \end{figure*}

   In the Figures \ref{VLT1} and \ref{VLT2} we show 3--colour--composites derived from the {\sl VLT} 
   J--H--K$_{\mathrm s}$ data, whereby Figure \ref{VLT2} shows the inner part of Figure \ref{VLT1} 
   in more detail and serves as a reference for the radio observation nomenclature we have adopted to
   indicate the various components.\\
   Figure \ref{VLT1} displays the global morphology of the star--forming
   complex G9.62+0.19. The extended diffuse region in the centre of the image 
   dominates the near--infrared emission and can be related to the more evolved
   H{\small II} component B in Fig. \ref{VLT2}. This region harbours a cluster of
   modestly embedded stars. Many of them show infrared excess. Recently, Persi et al. (\cite{Persi}) analysed 
   in detail this population of stars in the G9.62+0.19 B and C regions.
   
   Eastwards of this emission, one can notice an extinction gradient. Hence, 
   the neighbouring area, beginning a few arcsec away from the eastern border 
   of component B, appears darker and more obscured. The
   superimposed {\sl MSX} contour lines which we derived from the  
   8.28 $\mu$m image of the related {\sl MSX} source 
   (Egan et al. \cite{Egan1}) 
   experience the same decrease as the NIR emission towards the eastern darker region. 
   This is a hint for a similar behaviour of the optical depth in the near-- and mid--infrared.
   In fact, we categorise this eastern region as a so--called Infrared Dark Cloud
   (IRDC -- see Egan et al. \cite{Egan2}).
   
   The transition stripe between the two above--mentioned regions
   is aligned roughly along the north--south direction and is populated with
   several more compact radio sources named C, D, E and F (Hofner et al. \cite{Hofner2}, see
   also Fig.~\ref{VLT2} for reference).
    
   Mainly in the southwest of the radio component C we detect strong K band emission. The peak
   positions of the 3.6 cm and the NIR emission are clearly separated by almost 3 arcsec. 
   This shift is obvious in Fig. \ref{VLT2},
   it is real and not an effect of an unprecise astrometry. The
   K narrow-band data 
   and polarimetry data (Secs. \ref{Knarrow data} and \ref{pol data}) 
   will enlighten these circumstances.   
   
   The components D and E are ultracompact 
   H{\small II} regions (Hofner et al. \cite{Hofner2}, Testi et al. \cite{Testi2}), whereby E might be associated either
   with an expanding ionised shell or with a wide--angle molecular outflow (Minier et al. \cite{Minier}). 
   Especially source D has a strong peak at 3.6 cm which can be recognized via the superimposed
   {\sl VLA} contours in Fig.~\ref{VLT2}. Both components show no counterparts in our NIR data. 
   We will discuss the consequences of these findings in Sect.~\ref{why}.

\begin{figure*}
   \centering
     \includegraphics[width=\textwidth]{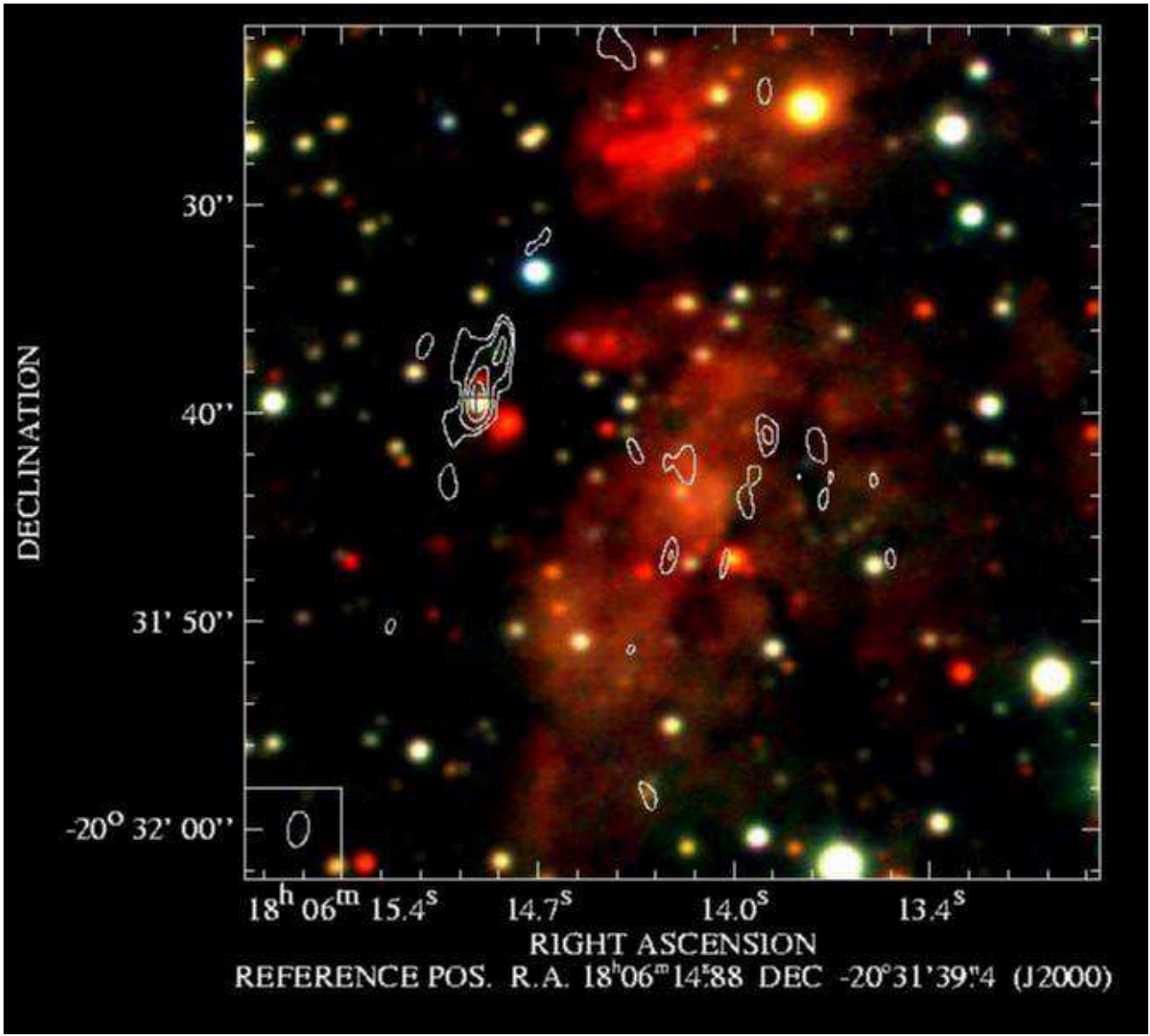}
      \caption{The same field of view as in Fig. \ref{VLT2}, but now the overlaid contour lines trace the thermal
               NH$_3$(5,5) emission of the HMC as measured by Hofner et al. (\cite{Hofner1}). The ellipse in the 
	       lower left corner indicates the size of the synthesized VLA beam. The black cross at 
	       the reference position marks the faint peak of the 3.6-cm HMC emission (Testi et al. \cite{Testi2}).
              }
         \label{VLT_NH3}
   \end{figure*}

\begin{figure*}
\centering  
\vspace{0cm}
\includegraphics[width=\textwidth]{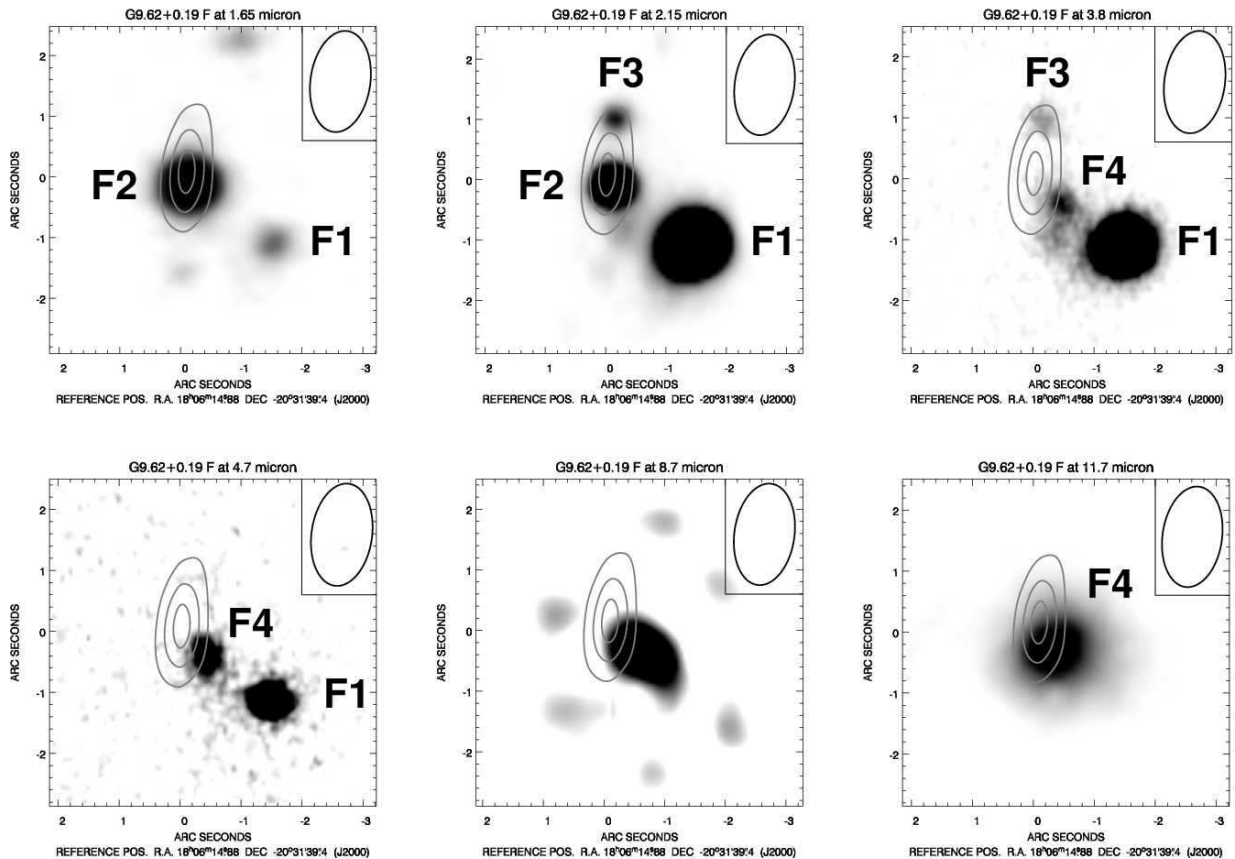}
\vspace{0.5cm}
      \caption{ Cutout gallery showing the four objects of the HMC region 
               G9.62+0.19--F. Upper row from left to right: H, K$_{\mathrm s}$,
               L' band. Lower row from left to right: nb\_M, N1, 11.7 $\mu$m. 
	       The size of the cutout box is ca.
	       $5.\!\!''4 \times 5.\!\!''4$.
               While the foreground star (F2) quickly fades towards 
	       longer wavelengths, another object very close to it (F4) gets stronger and finally. 
	       dominates the region. We presume that it 
	       is a potential trace of the HMC whose location is here indicated by the upper
	       contours of the NH$_3$(5,5) emission . While all the other images
	       are taken from the plain imaging data, the N1 image at 8.7 $\mu$m 
	       shows the TIMMI2 data after deconvolution with the
	       standard star HD169916. Note that the smaller spots in that particular image
	       are deconvolution artefacts that change their position and strength when
	       varying the deconvolution parameters.
               }\label{gallery}
   \end{figure*}

   The HMC component F was originally defined as the peak in NH$_3$(5,5) and NH$_3$(4,4) transition 
   measurements, conducted with the VLA (Hofner et al. \cite{Hofner1}, Cesaroni et al. \cite{Cesaroni1}). 
   The black plus sign in Fig.~\ref{VLT2} pinpoints the peak position of this molecular line emission.
   (In Fig. \ref{VLT_NH3} we overlay the contours of the HMC NH$_3$(5,5) emission onto the NIR data
   for comparison.)
   Further radio--interferometric
   observations could reveal emission peaks in the 2.7 mm dust continuum (Hofner et al. \cite{Hofner2}) as well as
   in the cm free--free continuum (Testi et al. \cite{Testi2}) at the position of the HMC component F.
   Testi et al. (\cite{Testi1}) have reported a K--band detection at the coordinates of the HMC. 
   Our infrared imaging reveals an intriguing structure in this HMC region which one can decompose
   into at least three distinct objects. We refer to Fig.~\ref{gallery} for the numbering we have chosen for
   the objects and which will be used in this paper.
   On the basis of our high--resolution {\sl VLT} data (pixel scale
   = 0$.\!\!''$15/px, seeing 0$.\!\!''$55) one can presume that this previous K--band observation of
   Testi
   (pixel scale = 0$.\!\!''92$/px) was suffering from the lower 
   resolution and confusion among the objects in that line of sight which can now 
   clearly be separated. We refer to our appendix \ref{astrometry} where we discuss in detail our
   approaches to derive an accurate astrometry for our data. \\
   Near the HMC peak coordinates, we find a ``red'' compact source according to the 3--colour--composite
   (object F1 in Fig.~\ref{gallery}). However, F1 does not coincide with the hot core peak F, but is displaced
   by 1.7 arcsecs.
    Although this finding suggests that the compact NIR emission F1 is not a direct trace
   of the innermost HMC region but a distinct object we should note that according to Cesaroni et al. 
   (\cite{Cesaroni1})
   the warm ammonia emission is extended over several arcsec and a source size of 3 -- 5
   arcsec was derived. Since 1 arcsec translates to 0.028 pc (assuming a distance of 5.7 kpc) 
   F1 might still be affected by warm and dense molecular gas of the actual HMC.
   Thus, we have derived the K$_{\mathrm s}$ band magnitude of object F1 by using
   S234--E (K$_{\mathrm s} = 12.070$ mag) and S071--D (K$_{\mathrm s} = 11.839$ mag) (Persson et al. 
   \cite{Persson}) as photometric standards. 
   By performing a PSF photometry on the mosaiced science frame we derived a count rate for
   our object. To be in accordance with 
   the 10$''$ aperture photometry applied by Persson et al. (\cite{Persson}) we also used this method 
   (in form of the {\sl MIDAS}\footnote{Munich Image Data Analysis System} {\sc magnitude/circle} routine) to
   derive the count rates for the standard stars. Corresponding to this
   procedure the F1 magnitude is K$_{\mathrm s} = 13.6$ mag with 0.1 mag as a conservative error limit. 
   Testi et al. (\cite{Testi1}) had derived K$= 12.9 \pm 0.08$ mag, probably for all the 3 objects visible
   at  2.16 $\mu$m in the F region in Fig.~\ref{gallery}.  \\  
   At the position of the HMC 3.6~cm radio peak we find another much fainter
   compact source (object F2 in Fig. \ref{gallery}). In order to evaluate this fact one should remember 
   that this area near the IRDC is in general
   very extincted. It is possible to derive a preliminary extinction estimation for this region 
   from the data contained in Schlegel et al. (\cite{Schlegel}) which results in A$_\lambda$ values 
   between 31.2 mag (J band) and 12.7 mag (K band). Therefore, other objects in the F region appear to be
   strongly reddended, with (H--K) values clearly exceeding 1.5 mag.
   However, this object has a ``yellow'' colour index with (H--K) being only 0.72 mag. By inserting the data
   for F2 into a JHKs two--colour diagram we see that the F2 colours can be explained almost completely by 
   interstellar reddening. Therefore, we conclude
   that F2 is a foreground star, not physically related to the star--forming region. A definitive answer cannot be given
   until high-resolution NIR spectroscopy is performed with an instrument capable of resolving the
   objects in the HMC region (e.g., NACO at the VLT).\\
   A third very faint source (object F3) is visible less than 1$''$ to the north of the foreground star. This object
   appears very red because it is only detected in the K$_{\mathrm s}$ band and not in the J or H band.
   An interesting fact to mention is that, according to our astrometry, an H$_2$O maser (the first
   one in the H$_2$O maser list in Tab. \ref{pos}) is only 0.25 arcsec apart from this object's 
   position. \\
   The photometry of the objects in the HMC region is summarised in Table \ref{mag}.

\begin{center}

  \begin{table*}[htb]
      \caption[]{Compilation of positions for features in the HMC region}
         \label{pos}

 \hspace*{3cm}        \begin{tabular}{llll}        
            \hline
	    \hline
            \noalign{\smallskip}
            Feature         &  RA(J2000)   & DEC(J2000)     & Telescope \\
            \noalign{\smallskip}
            \hline
            \noalign{\smallskip}
            NH$_3$(4,4)              & 18 06 14.85  &  -20 31 38.91  &  VLA -- D$^{\mathrm{(1)}}$ \\
            NH$_3$(4,4)              & 18 06 14.81  &  -20 31 39.41  &  VLA -- C$^{\mathrm{(1)}}$ \\
            NH$_3$(5,5)              & 18 06 14.87  &  -20 31 39.41  &  VLA -- C$^{\mathrm{(2)}}$ \\
            CH$_3$CN(6-5)            & 18 06 14.89  &  -20 31 39.20  &  OVRO$^{\mathrm{(3)}}$ \\ 
            2.7~mm cont.             & 18 06 14.89  &  -20 31 38.90  &  OVRO$^{\mathrm{(3)}}$ \\
            3.6~cm cont.             & 18 06 14.88  &  -20 31 39.37  &  VLA -- BnA$^{\mathrm{(4)}}$ \\
            H$_2$O masers            & 18 06 14.88  &  -20 31 38.39  &  VLA -- B$^{\mathrm{(5)}}$ \\
                                     & 18 06 14.87  &  -20 31 41.21  & \\
                                     & 18 06 14.81  &  -20 31 37.43  & \\
                                     & 18 06 14.93  &  -20 31 41.55  & \\
                                     & 18 06 14.86  &  -20 31 40.59  & \\
                                     & 18 06 14.78  &  -20 31 34.71  & \\
                                     & 18 06 14.87  &  -20 31 37.57  & \\
                                     & 18 06 14.67  &  -20 31 31.44  & \\
                                     & 18 06 14.98  &  -20 31 42.53  & \\
            \noalign{\smallskip}
            \hline
            \noalign{\smallskip}          
            object F1                & 18 06 14.78  & -20 31 40.48   & ISAAC$^{\mathrm{(6)}}$ \\
            object F2 (K$_{\rm s}$)  & 18 06 14.87  & -20 31 39.50   & ISAAC$^{\mathrm{(6)}}$ \\
            object F3                & 18 06 14.87  & -20 31 38.39   & ISAAC$^{\mathrm{(6)}}$ \\
            object F4 (nb\_M)        & 18 06 14.85  & -20 31 39.70   & ISAAC$^{\mathrm{(6)}}$ \\

            \noalign{\smallskip}
            \hline
         \end{tabular} \\
 
\begin{list}{}{}
\item Ref.: (1) Cesaroni et al. (\cite{Cesaroni1}) ,  (2) Hofner et al. (\cite{Hofner1}) , \\ 
            (3) Hofner et al. (\cite{Hofner2})   ,  (4) Testi et al. (\cite{Testi2})  , \\
            (5) Hofner \& Churchwell (\cite{Hofner3})   ,  (6) this article
\end{list}
   \end{table*}

\end{center}

 \subsection{L' and nb\_M band data}\label{L data}

   This wavelength range is interesting because it fills the gap between the K$_{\mathrm s}$ 
   and the N band. Naturally, the signals of most of the NIR--blue field stars strongly fade and 
   finally disappear. Hence, they do not cause confusion anymore. Still, enough objects remain in the 
   FOV of {\sl ISAAC} to perform an accurate astrometry.
   Hereby we want to direct the attention to an interesting detail
   (see Fig. \ref{gallery}). \\
   In the K$_{\mathrm s}$ band image one can clearly see the presence of three objects in the HMC 
   region F. 
   The ``red'' object F1, already mentioned in subsection \ref{NIR data}, is dominating. The object F2
   which we estimated to be an unrelated foreground star is almost 1.7 arcsec apart in the north--east
   direction. Above of this we detect the third very red and weak source F3. Please note that the 
   location of the foreground star is underlaid with weak and diffuse emission.\\
   When we now switch to the L' band we notice that the point--like star F2 in the middle has almost
   totally disappeared.  Instead, we now clearly see emission arising from a slightly different location,
   that reveals the presence of another object F4.
   This is evident, when we compare the relative positions of the emission centres.
   We mentioned the distance of 1.7 arcsec between the point--like star F2 and
   the ``red'' object F1 in
   the K$_{\mathrm s}$ band. The centre of F4 in the L' band image is clearly shifted
   compared to the F2 position; it is now only 1.3 arcsec away from object F1.
   This shift is persistent also in the 4.07 $\mu$m and the nb\_M band image; it is a small
   effect, but noticeable.  In the following, the nature of object F4 will become more clear.\\

 \subsection{N and Q band data}\label{N data}

\begin{figure}
   \centering
   \includegraphics[width=12.75cm]{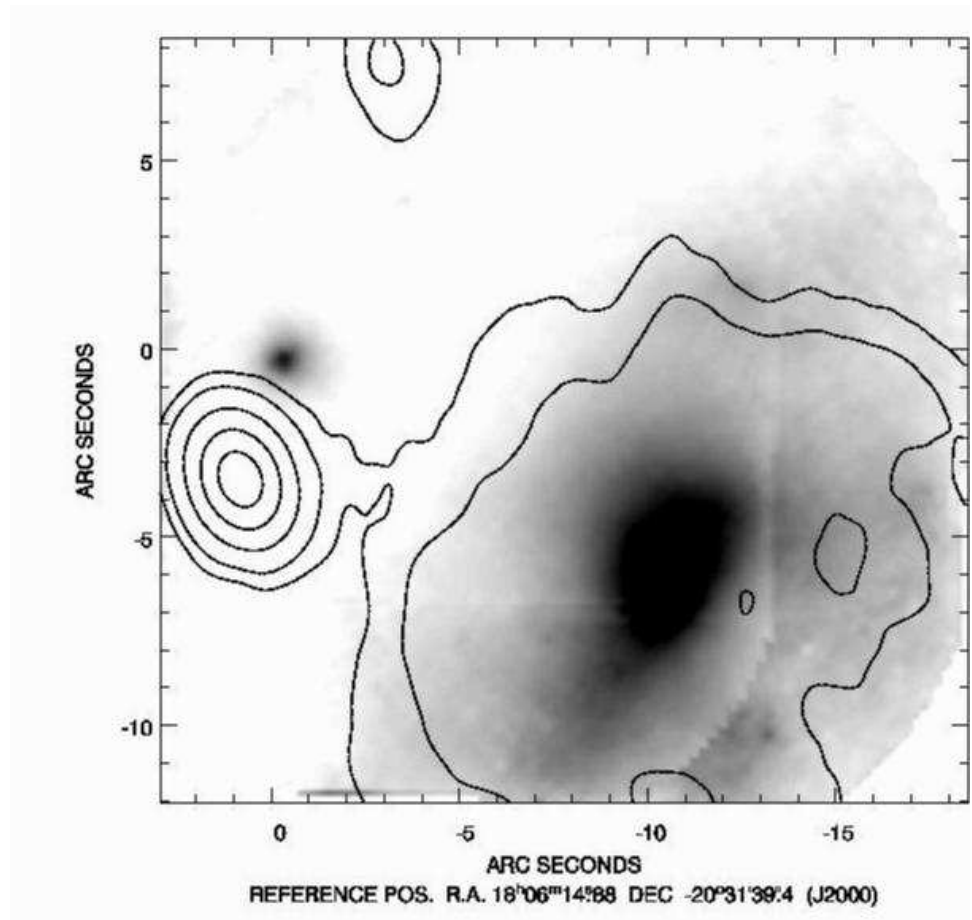}
      \caption{ Inverse gray--scale representation of the 11.7 $\mu$m SpectroCam-10 image displaying 
      a smaller cutout of the G9.62+0.19 region. The weaker but nevertheless clearly visible spot to the 
  east represents emission from the region of the HMC component F and arises probably  from object F4 (see
  also Fig. \ref{gallery}). 
  Overlaid are the 1.3 cm continuum {\sl VLA} contours from Cesaroni et al. (\cite{Cesaroni1}). The 
  compact radio source to the south--east of the compact MIR emission is the UCH{\sc ii} region G9.62+0.19 D.
              }
         \label{SC10}
   \end{figure}

   Fig. \ref{SC10} shows the central part of Fig. \ref{VLT1} but now at $\lambda$ = 11.7 $\mu$m
   obtained with {\sl SpectroCam-10}. To amplify the morphological structure,
   we applied 25 iterations of a modified Maximum Entropy Method  for noise suppression, 
   based on the wavelet transform and developed by Pantin \& Starck (\cite{Pantin}). The extended 
   emission, coming from component B, dominates also in the mid--infrared. 
   A second, but compact source is found near the extended B component.  
    The accurate position and hence the correct
   interpretation of this compact source has been a matter of debate during recent years (cf. De Buizer et al.
   \cite{Buizer}, \cite{Buizer4}, Stecklum et al. \cite{Stecklum1}, Persi et al. \cite{Persi}). 
   In Appendix \ref{astrometry2} we describe the circumstances and explain the result of our new astrometry.
   We find that the compact MIR emission is coming from the HMC region, contrary to the results of 
   De Buizer et al. (\cite{Buizer4}). We identify it with object F4 (see Fig. \ref{gallery}), i.e., with
   the HMC conterpart.\footnote{Persi et al. (\cite{Persi}) also identified the MIR source with the HMC on the
   basis of low signal--to--noise data and related it to the K band emission found by Testi et al. (\cite{Testi1}). 
   However, they apparently underestimated the significance of the multiplicity of objects in the G9.62+0.19--F 
   region and thus could not distinguish between F1 (dominating in the K band) and F4 (dominating in the N band).}   
   For a determination of the flux coming from the compact source F4, we observed the star
   $\delta$ Oph just before we imaged the G9.62+0.19 region. This isolated star is a strong {\sl IRAS}
   source with a 12 $\mu$m flux of 149.7 Jy. By applying an aperture photometry to the compact source 
   (using a 3 arcsec diaphragm) and the photometric reference star, we derived a flux of 
   (0.60 $\pm$ 0.10) Jy for F4. \\
     The 8.7 $\mu$m N1 data proved their value especially regarding the MIR astrometry, as shown in 
   Appendix \ref{astrometry2}. We show these data in Fig. \ref{TIMMI2}. To produce this image
   the matrix restoration method described in appendix \ref{Landweber} was applied to the N1 band 
   TIMMI2 data. Thereafter, the restored data were overlaid as contours on the NIR colour composite. 
   As can be easily seen, they comprise a much
   larger field of view than the {\sl SpectroCam-10} data in Fig. \ref{SC10}.
   We also derived the flux of the compact MIR emission at this wavelength of 8.7 $\mu$m.
   As is evident in Fig. \ref{TIMMI2} and Fig. \ref{gallery} the emission is elongated; most of the flux
   stems from object F4, but object F1 might still contribute a small fraction to the total signal. 
   We did not try to disentangle F4 and F1 and only measured the total flux with an adequately sized 
   aperture for the photometry. One should keep in mind that the N1 filter covers a prominent
   PAH feature, the related emission also peaks around  8.6 -- 8.7 $\mu$m (e.g., Peeters et al.
   \cite{Peeters2}). In fact, such PAH emission seems to be a common feature of the majority of
   compact H {\sc II} regions and their surroundings (Peeters et al. \cite{Peeters1}).\footnote{ We mention our experience with another massive star--forming 
   region (Apai et al. \cite{Apai}) where PAH emission (also covered by the MSX A band) clearly 
   affects the SED of the central object. } 
   Thus, the measured N1 flux is 
   probably affected by additional contributions from those small PAHs that are suscptible to
   transient heating and often attain a higher (non--equilibrium) temperature than the larger normal  
   dust grains.  Indeed, the N1 flux of (1.10 $\pm$ 0.10) Jy is higher than 
   the 11.7 $\mu$m flux of F4. \\  
   Although the atmospheric conditions during the Q2 band observations at 18.75 $\mu$m were not photometric
   we tentatively report the measured flux of (22.8 $\pm$ 7.0) Jy for the compact emission blob we 
   assume to be identical  to F4 -- no other reference objects appeared in the data except of the
   extended emission of component B. The quoted relatively high uncertainty of 30 \%  is mainly caused
   by the non--perfect sky. \\
   All the MIR measurements are also included in the SED plot of object F4 in Fig. \ref{SED} and draw an
   enlightening picture of the energetics of the HMC. In Sect. \ref{RT} we will further elaborate on the 
   consequences of our findings for object F4.

   \begin{table*}
      \caption[]{ Measured IR fluxes in Jy (corresponding magnitudes in parenthesis) for the objects in the HMC region}
         \label{mag}
      \renewcommand{\arraystretch}{1.2}
         \begin{tabular}{ccccc}
            \hline
	    \hline
            \noalign{\smallskip}
            Band name       & object   & object   & object & object  \\
                            &  F1      &  F2      &  F3    &  F4     \\
            \noalign{\smallskip}
            \hline
            \noalign{\smallskip}
            J               &  --                 & $2.21\times10^{-4}$ (17.15)    & --                & --                 \\
            H               & $6.19\times10^{-5}$ (18.00)    & $5.25\times10^{-4}$ (15.72)    & --                & --                 \\
            K$_{\mathrm s}$ & $2.25\times10^{-3}$ (13.60)    & $6.57\times10^{-4}$ (15.00)    & $1.67\times10^{-4}$ (16.49)  & --                 \\
            L'              & $2.43\times10^{-2}$ (10.04)    & --                  & $1.00\times10^{-3}$ (13.50)  & $2.95\times10^{-3}$ (12.33)   \\  
            nb\_M           & $3.10\times10^{-2}$ ( 9.21)    & --                  & --                & $1.49\times10^{-2}$ (10.01)   \\
	    N1              & $1.10\            $ ( 3.90)$^{\rm{a}}$    & --        & --                & $1.10$ ( 3.90)$^{\rm{a}}$ \\
            N               & --                  & --                  & --                & $6.00\times10^{-1}$ ( 4.55)   \\
	    Q2              & --                  & --                  & --                & $2.28\times10^{1}$ (--0.85)$^{\rm{b}}$   \\
            \noalign{\smallskip}
            \hline
         \end{tabular}
     
\begin{list}{}{}
\item[$^{\mathrm{a}}$] { In the undeconvolved N1 band, the objects F1 and F4 
                        merge to one entity, whose combined flux is
			reported.}
\item[$^{\mathrm{b}}$] { Note that the Q2 flux is affected by a relatively large uncertainty of around 30\% due to 
                           non--photometric sky conditions (see text).}
\end{list}
   \end{table*}
   
  \subsection{H$_2$ and Br$\,\gamma$ narrow band imaging data 
  }\label{Knarrow data}
  
   In Fig \ref{Brg_H2} we combined the Br $\gamma$ data (blue channel) and the H$_2$ data 
   (red channel). The average of the signal of both filters was put into the green channel. 
   The Br $\gamma$ emission of the compact H{\sc ii} region B clearly dominates the overall
   appearance of the star--forming complex. The ultra-- and hypercompact H{\sc ii} regions
   D and E are not visible at all due to the large K band extinction towards these components
   (see section \ref{why} for a discussion). More interesting, the signal arising from the
   region of the radio C component appears almost white in this colour composite; thus, 
   it represents just the 
   contribution of the K band continuum without strong H$_2$ or Br $\gamma$ line emission.
   The same we find for the NIR objects around the HMC region F.
   
   The H$_2$ data reveal a compact emission region located between the components B and C.
   As a whole, it appears as a flat ellipsoid with a large axis of about 5$''$, but it
   is dissected into several emission maxima and minima.  This feature was also mentioned by Persi et al.
   (\cite{Persi}) and we report here the positions of the two main components according to our 
   astrometry.\footnote{ For the probable cause of the astrometric disagreement between our positions and
   the one of Persi et al. (\cite{Persi}) we refer to the end of Appendix \ref{astrometry}.} Component 1:
   $\alpha_{2000} = 18^{\rm h} 06^{\rm m} 14 \fsecs42, \delta_{2000} = -20^{\circ} 31' 36 \farcs8 $, 
   Component 2:    $\alpha_{2000} = 18^{\rm h} 06^{\rm m} 14 \fsecs34, \delta_{2000} = -20^{\circ} 31' 36 \farcs8 $. 
   The projected distance to the hot core region F
   is roughly 7 arcsec (i.e. 0.2 pc if assuming $d = 5.7 \,$kpc). In principle, such a distance
   would allow an association with the molecular outflow arising from component F (Hofner et al. \cite{Hofner4}).
   One problem is that the outflow is well aligned along the
   line--of--sight. Thus, if the interpretation of shock excitation due to interaction with the
   outflow holds, then the outflow driving source and the H$_2$ emission region cannot be located
   in the same plane of the sky, implying that the latter might be detached from the actual 
   star--forming complex G9.62+0.19. Another possibility is the existence of another outflow in
   this region that was not detected in our previous studies (Hofner et al. \cite{Hofner2}, \cite{Hofner4}).
   An alternative consideration would be that a considerable fraction (if not all) of the
   detected H$_2$ emission is in fact fluorescence, excited by UV photons of the 
   nearby H{\sc ii} regions (e.g., Black \& van Dishoeck \cite{Black} and Draine \& Bertoldi \cite{Draine2} 
   for a theoretical treatment,
   e.g., Fernandes et al. \cite{Fernandes} and McCartney et al. \cite{McCartney} for observational 
   evidence in the case of other astronomical
   objects). The components B or C could act as UV photon donor. Only further spectral 
   analysis, comprising several H$_2$ roto--vibrational transitions, can distinguish between
   the collisional shock model and the fluorescence model.   However, we mention that the H$_2$ feature
   is not a pure emission line object, we also see diffuse continuum emission from this position 
   in the L' and nb\_M band, as well as in the 2.09 $\mu$m filter used for the polarimetric imaging (see the
   directly following paragraph).\\
   
\subsection{K narrow band imaging polarimetry 
  }\label{pol data}  
    
   The polarimetric data were combined in order to generate a map of the linear polarisation of G9.62+0.19.
   We  chose the standard approach for obtaining the polarimetric quantities, using the four
   intensities measured with the Wollaston prism (see Sect. \ref{pol_obs}). The Stokes parameters can 
   be defined as:
   \mylabel{Stokes-all}{-1.4cm}
   \alpheqn
   \begin{eqnarray}
      Q & = & I(0^{\circ}) - I(90^{\circ}) \; , \\                                   \label{Stokes-a}
      U & = & I(45^{\circ}) - I(135^{\circ}) \; ,  \\                                \label{Stokes-b}
      I & = & (I(0^{\circ}) + I(90^{\circ}) + I(45^{\circ}) + I(135^{\circ}))/2 \; . \label{Stokes-c}
    \end{eqnarray}
    \reseteqn
    The degree of linear polarisation ($p_{\mathrm{lin}}$) and the respective polarisation angle
    ($\Theta_{\mathrm{lin}}$) can then be derived by using:
    \begin{equation}\label{p_lin}
      p_{\mathrm{lin}} = \frac{\sqrt{Q^2 + U^2}}{I} \; \; \; \; \; \; \; \; \; \; \; \;
      \Theta_{\mathrm{lin}} = \frac{1}{2} \, \mathrm{arctan} \, \frac{U}{Q}  \; .
    \end{equation}
   In Fig. \ref{VLT4} we show the resulting polarisation map superimposed on the Stokes--$I$ image (see
   equation \ref{Stokes-c}) derived from the very same data set.  We mention that the degree of
   linear polarisation as derived from Eq. (\ref{p_lin}) poses an upper limit for the true value
   of $p_{\rm lin}$ because of the non--standard error distribution of this quantity (e.g., Simmons \&
   Stewart \cite{Simmons}). Thus, before plotting we debiased $p_{\rm lin}$ using the approach given in
   the appendix to the paper of Wardle \& Kronberg (\cite{Wardle}).\\
   NIR imaging polarimetry of circumstellar matter proved to be a valuable tool for revealing the illuminating 
   source, even in cases where it is hidden from the direct view due to enhanced extinction 
   (e.g., Tamura et al. \cite{Tamura}, Yao et al. \cite{Yao2}, Stecklum et al. \cite{Stecklum3}). 
   In the  north--west of Fig. \ref{VLT4}, we find  two adjacent regions of strongly enhanced 
   linear polarisation, covering the positions of component C (cf. Fig. \ref{VLT2}) and of the H$_2$ emission
   feature (cf. Fig. \ref{Brg_H2}). The  debiased polarisation degree partly exceeds 50 \%, indicative of 
   single--scattering events.  At first glance, the arrangement of the polarisation vectors implies a common illuminator for
   both regions which seems to be located south--east of them towards the more obscured region of the
   neighbouring IRDC (see Fig. \ref{VLT1}).  The ellipse in Fig. \ref{VLT4} indicates the 
   1$\sigma$ confidence region where the 
   illuminator is probably located. It is derived by tracing the intersections of lines perpendicular to the
   polarisation vectors whereby we only took the vectors with $p_{\rm lin} > 20$ \%  into account. 
   The most probable location of the illuminator is the centre of mass (CoM) of the intersection
   points, whereas their distribution yields an error estimate. A 2--dimensional Gaussian is fitted to
   the distribution of intersection points, the computed $\sigma$ parameters, which define the 
   fitted Gaussian,  are used as axes of the plotted ellipse (Weintraub \& Kastner 
   \cite{Weintraub}, Stecklum et al. \cite{Stecklum3}). The ellipse is quite eccentric because the available 
   polarisation vectors reside in only 2 of the 4 image quadrants. When looking carefully at the
   polarisation pattern one sees that the polarisation vectors related to the H$_2$ feature are more or less
   oriented north--south. Hence, when judging by eye, one would expect the centre of mass of all intersection points to be
   located more to the south--east of the plotted ellipse. An explanation for the different location in
   the plot (Fig. \ref{VLT4}) is that we can use some 180
   polarisation vectors near region C for the CoM estimation. But only some 30 vectors from the H$_2$ feature region can be
   taken into account that fulfill the conditions of sufficiently high degree of polarisation and
   acceptable signal--to--noise ratio in the intensity image. Thus, the 
   polarisation vectors around component C will contribute far more intersection points than the vectors 
   of the H$_2$ feature and hence will simply "preponderate" the CoM estimation. Is this a hint for
   several illuminating sources? We mention that 
   the error $\Delta p$ is still around
   8 \% in the high polarised regions. For ratios $p/\Delta p > 3$ the typical error for the position
   angle can be approximated by $28.65^{\circ} \times \, \Delta p/p$ (Serkowski \cite{Serkowski}). This
   results in an error of $4^{\circ} - 9^{\circ}$ in our case. But this is of course a statistical error. It 
   could not explain the coherent rotation of a whole polarisation pattern in one part of the image 
   (i.e., the H$_2$ feature region). 
   The bottom line of these considerations is: The observed polarisation map does not fully 
   correspond to one coherent circularly symmetric pattern as expected in the case of only
   one illuminating object, thus indicating the possibility of a second illuminator.  \\
   The radio object E (Fig \ref{VLT2}) is situated inside the ellipse in Fig.
   \ref{VLT4} and is an illuminator candidate, at least for the scattered light near 
   component C. However, the position of E does not really fit
   with respect to the polarisation pattern of the H$_2$ feature.
   We mention that the hypercompact radio components  G, H, and I (Fig.
   \ref{VLT2}), whose nature is not yet clarified, are located very near the 1$\sigma$ 
   confidence region for the illuminator position and their respective positions are in better 
   agreement with the polarisation pattern of the H$_2$ feature. Thus, one of these sources might be
   responsible at least for the illumination of the H$_2$ feature.  
   Since we do not see clearly corresponding NIR counterparts for these objects, they are probably deeply
   embedded. Should one of these objects turn out to be an illuminator, this would have implications for
   the distribution of its circumstellar material. A pronounced asymmetry  of the ambient matter 
   distribution would be necessary to explain
   how the object is able to illuminate the highly polarised regions in the K band, while we do not 
   detect any K band emission from it in our line of sight. \\
   The K narrow--band and polarisation 
   data of component C prove their importance also regarding the general
   astrometry of the region. In Fig. \ref{VLT2} we see that for component C the overlaid VLA contours 
   of the UCH{\sc ii} region are shifted from the NIR counterpart. One could suspect an error in the 
   overall astrometry, if one interprets this K band emission either 
   to arise from Br $\gamma$ emission of the ionised gas 
    and/or as thermal emission from hot dust, which usually is well mixed with the ionised gas of the 
    UCH{\sc ii} region. 
   Our data show that the Br $\gamma$ fraction of the K band signal of component C is marginal 
   and that the dominant part of this radiation probably consists of scattered light. 
   Under these circumstances, there is no reason that the cm continuum contours match the near--infrared emission.



\section{Discussion}

  \subsection{What is the nature of the IR objects in the HMC region?}
  
   In the previous sections we could establish the newly discovered object F4 as the actual HMC infrared
   counterpart by means of thermal and mid--infrared observations and a careful astrometry. In this section
   we discuss the consequences that arise from this finding. Some more theoretical consideration might help
   to support our view. 
  
 \subsubsection{Radiative transfer computations}\label{RT}

    To understand the properties and unusual features of the G9.62+0.19--F hot core we performed some 
   basic continuum radiative transfer computations, utilising two selfconsistent radiative transfer codes
   (Manske \& Henning \cite{Manske}, Wolf et al. \cite{Wolf}). For this paper we will only consider
   spherically symmetric models, in accordance with the standard description of hot molecular cores 
   (e.g., Osorio et al. \cite{Osorio}), just to demonstrate the limitations of this concept in the case
   of our HMC. 
   As input parameters we used data from Hofner et al. (\cite{Hofner2}). They had 
   limited the spectral type of the central source to be B0.5 -- B1. The luminosity was estimated to be 
   $\approx 1.8 \times 10^4$ L$_{\odot}$  (see also Cesaroni et al. \cite{Cesaroni1}). The mass is not well constrained by
   other authors, but ranges between 55 and 160 M$_{\odot}$ (Hofner et al. \cite{Hofner2}). Several power laws for the 
   radial density gradient were applied ($\rho \sim r^{-\alpha}$ with $\alpha =
   0.0 \, . \,  . \, 2.0$). As grain material 
   a mixture of silicates (Dorschner et al. \cite{Dorschner}) and carbonaceous materials 
   (Preibisch et al. \cite{Preibisch}) was used.  The computed
   fluxes comprise the range of $10^{-8} - 10^{-4}$ Jy for $\lambda = 2.2 \, \mu$m
   and $10^{-4} - 5 \times 10^{-1}$ Jy for $\lambda = 11.7 \, \mu$m, respectively, and thus
   are at or below the detection limit of {\sl ISAAC}, while the situation for {\sl SpectroCam-10 \ } at 11.7
   $\mu$m is more relaxed. While parameter combinations could be found that more or less fit
   the observational data points for F4 beyond 10 $\mu$m, it was not possible to also fit the data up to 5 $\mu$m. \\
   In Fig. \ref{SED} we show the result of such a  radiative transfer computation
   and compare it to the measured fluxes of  the objects F1 and F4 mentioned in Sects. \ref{NIR data},
   \ref{L data}, and \ref{N data}. 
   The particular parameter set used for Fig. \ref{SED} was chosen to produce preferably high fluxes in the 2 --5 $\mu$m range.
   However, the synthetic SED should still be in accordance with the data points at longer wavelengths.  For
   these points, the best fit was reached with a constant density distribution, an outer radius of
   2600 AU, a total luminosity of
   1.88 $\times 10^4$ L$_{\odot}$, and a dust mass of 0.94 M$_{\odot}$ inside the model space (see Fig. \ref{SED}). The optical depth for
   this particular configuration was $\tau_{\rm v} = 673$. 
   It becomes evident that the F4 emission in the 2 -- 5 $\mu$m range is
   still more than an order of magnitude stronger than predicted by the model. We interpret this as support for the
   idea that spherical symmetry is not applicable in the case of G9.62+0.19--F. 
   However, a comprehensive two--dimensional modelling of the observed fluxes
   would be beyond the scope of this paper. 

To guide the eye, we added in Fig. \ref{SED} a modified blackbody curve to the radiative model SED that also fits
quite well the data points at longer wavelengths. The effective temperature of such a modified blackbody would be 90 K, the
wavelength dependence of the opacity is decribed by $\kappa_{\mathrm m} \sim \lambda^{-1.0}$ in our case. However,
the significance of such a temperature estimation is questionable for the object class of HMCs. As shown, e.~g., 
by Yorke (\cite{Yorke-1}) and Yorke \& Shustov (\cite{Yorke0}), the concept of an effective temperature and 
an effective radius is not
well defined for very dense dusty envelopes around protostellar objects, they can strongly deviate from the
blackbody or greybody nature. Furthermore, we should note that such an effective temperature might be
misleading in the following regard. It was shown by means of simple radiative transfer considerations,
e.~g., by Schreyer et al. (\cite{Schreyer0}) that the mass-averaged dust temperature, i.~e. the temperature
averaged over density and particle size distribution, is always lower ($\Delta T = 15 .. 30$ K) than the theoretical colour
temperature derived from the SED in the case of dense shells around massive YSOs. Thus, most of the mass in such a
shell will attain a clearly lower temperature than insinuated by the colour temperature. This is important when using
a temperature in order to derive dust masses from infrared or mm observations.

 The estimated luminosity of $1.88 \times 10^4 $ L$_{\odot}$ from our simple model is in good 
agreement with previous predictions (Cesaroni et al. \cite{Cesaroni1}). 
When we also take into account the measured excess flux of F4 at shorter wavelengths
(in comparison to the radiative transfer models and the modified blackbody approximation) 
we gain another 1000 L$_{\odot}$.
Hence, the entire luminosity is approximately 20~000  L$_{\odot}$.
This seems to prove that the HMC in G9.62+0.19--F 
indeed harbours a young high--mass star. However, caution is advisable, as we see in the case of the Orion 
Hot Core, where on a much finer scale than in our case several infrared sources can be distinguished 
(Dougados et al.\cite{Dougados}), and the discussion about the energy budget of the individual sources
has not settled yet (e.g., Gezari et al. \cite{Gezari1}, \cite{Gezari2}).\\


  \subsubsection{Hints from the polarimetry}\label{hints}
     
   The polarimetric data can provide additional information on the geometry of the
   observed regions. In a scenario where an outflow has opened the dense molecular shell of a
   molecular hot core, infrared radiation can escape through a channel of reduced density.  
   The light will be scattered at the cavity walls (``dust mirroring'') which represent an 
   increased column density of scatterers in the optically thin cavity environment 
   (Yao et al. \cite{Yao1}, \cite{Yao2}). These single scattering events 
   can lead to high degrees of linear polarisation (e.g., Fischer et al. \cite{Fischer}). Hence, we would 
   expect to perceive a clear increase of NIR polarisation.
   
   In contrast, the measured polarisation for object F1 is relatively low (10 $\pm $ 5) \%, especially 
   when it is compared to polarisation degrees in its immediate vicinity.\footnote{ It is difficult
   to confirm this claim just by looking at Fig. \ref{VLT4}. This kind of plot is well--suited to 
   visualise the polarisation pattern for extended emission but is often misleading around compact
   sources with steep intensity gradients. In these cases the polarisation has to be derived by 
   aperture photometry centred on the compact sources in all the polarimetric frames and then applying
   the Eqs. (\ref{Stokes-all}) and (\ref{p_lin}).} There, the polarisation 
   increases up to (33 $\pm $ 10) \% in the small region of diffuse emission to the southwest of F2, but not 
   including F2  or F1 (best to be seen in the 2.16 $\mu$m image of Fig. \ref{gallery} at the offset
   position (--0.3, --0.8). Based on the low--resolution NIR imaging of Testi et al. (\cite{Testi1}), we suggested in our 
   recent outflow paper (Hofner et al. \cite{Hofner4}) that object F1 itself represents infrared emission 
   arising from the HMC region which might be cleared in parts by the molecular outflow.
   Taking our high--resolution infrared data, the improved astrometry and the results of the polarimetry
   into account, we no longer believe that this model can satisfactorily explain the properties of 
   object F1. \\
   However, this scenario is certainly a good interpretation for
   the diffuse emission behind the foreground star. It appears
   unlikely that this diffuse emission is just the result of the reflection of light coming from the
   foreground star F2. In such a case of backscattering the reflected light should more closely
   resemble the colour behaviour of the illuminating star.  Moreover, there should be a tendency for a
   slighly more bluish SED of the backscattered light due to the decreasing scattering efficiency for
   increasing wavelengths. This is not observed. Instead, the diffuse emission is
   clearly redder than the star  and seems to be associated with object F4 whose emission strongly increases
   in the L' and nb\_M band, while the star F2
   strongly fades beyond 2.2 $\mu$m. We therefore think that this star F2 is clearly detached from the
   actual star--forming region. \\
   We should re--emphasise that both, the weak
   cm continuum emission and the NH$_3$(5,5) emission associated with the HMC component F,
   peak very close ($\le 0.\!\!''4$, cf. Table \ref{pos}) to F4 (see also Figs. \ref{VLT2} 
   and \ref{VLT_NH3}) which suggests an intrinsic entanglement between both components.
   Thus, we interpret F4 as arising from the inner parts of the 
   HMC outflow cavity. \\
    One detail is not yet clarified: Is the L' and nb\_M band signal of F4 caused by direct thermal
    dust emission of hot grains very near to the embedded power source of the HMC? Or can a
    considerable fraction of the overall emission be attributed to reprocessed light scattered on the
    cavity walls? In the latter case one has to assume the existence of larger dust grains in the
    HMC envelope to ensure a reasonable high scattering efficiency\footnote{Usually, the scattering efficiency
   is strongly diminished at wavelengths beyond $\approx 2 - 3 \, \mu$m when considering only dust
   with the standard MRN size distribution (Mathis et al. \cite{MRN}) typical for the
   interstellar medium, i.e., the overwhelming
   majority of the dust particles is sub--micrometer--sized with a typical diameter of
   $\approx 0.1 \, \mu$m. } for wavelengths of 3 -- 5 $\mu$m. \\
These ideas about larger grains are more than a wild speculation.
Grain growth is expected to occur in YSOs and their near vicinity, 
in particular, in circumstellar disks (e.g., Beckwith \& Sargent 
\cite{Beckwith1}, Beckwith, Henning, \& Nakagawa \cite{Beckwith2}, D'Alessio, Calvet, \&
Hartmann \cite{D'Alessio}). Recent observations
(e.g., McCabe et al. \cite{McCabe}, Shuping et al. \cite{Shuping}) seem to indicate --
cum grano salis -- the existence of a considerable amount of micron--sized particles
in disks around low--mass YSOs. Systematic investigations for the case of massive YSOs
seem to be more scarce, especially regarding HMCs. However, for the best--known region
of ongoing massive star formation, Orion, polarimetric observations at 2.2 and 3.8 $\mu$m 
point to the existence of larger grains in the Orion Molecular Cloud in general (Rouan \& 
Leger \cite{Rouan}) and in particular in the BN/KL region containing the Orion Hot Core 
(Minchin et al. \cite{Minchin}, Dougados et al. \cite{Dougados}). \\
Thus, a future task will be to conduct high--resolution polarimetric measurements in the K,
L', and nb\_M band. Adaptive optics systems like {\sl CONICA/NAOS} at the VLT seem to be 
most suitable to ensure high resolution (0 \farcs1) combined with high sensitivity -- 
characteristics that are absolutely necessary in order to deliver meaningful polarimetric 
results in the case of G9.62+0.19--F.

  \subsubsection{Ordinary stars within the HMC vicinity?}

   We have pointed out already, that object F2 very near the HMC is probably a
   foreground star. Is it also possible that object F1 is a normal star whose 
   photospheric emission we see? We have several reasons to decline this possibility. \\
   We start with the following estimate: We use the well-known correlation between
   the colours of a star and its colour excess due to additional reddening:
   \begin{equation}\label{colour}
       H - K \, = \, (\, H - K \,)_{\, 0} \, + \, E_{\, \mathrm{H-K}} \; .
   \end{equation}   
   Herein, ($H - K$)$_0$ is the unreddened intrinsic colour of the star and 
   the colour excess $E_{\, \mathrm{H-K}}$ can be expressed as
   \begin{equation}\label{excess}
       E_{\, \mathrm{H-K}} = A_{\mathrm H} - A_{\mathrm K} = 
       \Big( \frac{A_{\mathrm H}}{A_{\mathrm K}} - 1 \Big) \; A_{\mathrm K} \; .
   \end{equation}
   According to Mathis (\cite{Mathis}), the $A_{\mathrm H}/A_{\mathrm K}$ ratio for the standard
   interstellar reddening law ($R_{\mathrm V} = 3.1$) is around 5/3. We can use the
   intrinsic stellar colours (e.g., Wegner \cite{Wegner} for O and B stars,
   Bessell \& Brett \cite{Bessell} for stars of type A -- M) and our measured value $H - K = 4.4$ mag to
   derive the K band extinction $A_{\mathrm K}$. O stars of all luminosity classes
   have ($H - K$)$_0$ values between $-0.08$ mag and $-0.02$ mag. In combining (\ref{colour}) and
   (\ref{excess}) we find that $A_{\mathrm K}$ is always larger than 6.6 mag for our case.
   Together with the distance module of 13.8 (assuming $d = 5.7$ kpc) and the measured 
   K band magnitude of 13.6 mag (see Table \ref{mag}), the absolute K magnitude $M_{\mathrm K}$
   would be $- 6.8$ mag. Moreover, if we take into account that all O stars have 
   ($V - K$)$_0$ values between $-0.7$ mag and $-0.9$ mag, the absolute visible magnitude $M_{\mathrm V}$
   would be brighter than $-7.5 \,$ mag! O and B stars do not feature such  ``bright'' 
   $M_{\mathrm V}$ values, only very extreme supergiants (luminosity class 1a-0) would exhibit
   such a characteristic. Furthermore, if we considered late--type giants we would also have to
   deal with extreme luminosity classes which, in addition, would cause a severe dilemma
   to explain the existence of a very old star almost in the centre of a young star--forming 
   region.

   Striking evidence against photospheric emission comes from the K band medium--resolution spectrum 
   (Fig. \ref{spectrum}) which was taken with SOFI at the 3.5-m NTT on La Silla, Chile (Testi \cite{Testi3}). 
   The slit was 
   centred on the object F1, but the spectrum probably covers also minor contributions 
   from the foreground star F2. Around 1.87 $\mu$m the transmission of the atmosphere 
   strongly decreases due to strong H$_2$O absorption, herewith separating the near--infrared H and K 
   band. The strong feature around 1.87 $\mu$m probably arises from the fact that this atmospheric 
   imprint was not totally canceled out in the calibration. In the range between 1.95 and 2.40 $\mu$m 
   we see a spectrum without prominent features. Especially late--type stars with their relatively
   cool atmospheres should show many metallic lines and even molecular lines. A characteristic
   feature for such stars are more or less strong CO absorption features between 2.29 $\mu$m and 2.40 $\mu$m
   (Ramirez et al. \cite{Ramirez}, Bieging et al. \cite{Bieging}), but we only see 
   an almost featureless spectrum monotonically rising towards the K band edge, which is
   an imprint of continuum dust emission.

   These points speak clearly against the ``naked star'' interpretation for the object
   F1. They rather indicate the presence of an embedded object within a dusty shell. 
   In a way, this resembles the situation of the Orion BN/KL region where several IR objects are 
   located in the vicinity of the Orion Hot Core. 
   It might be tempting to see our object F1 as an analogue to the BN object with regard to
   the general appearance and the dusty shell nature of both objects. However, this analogy does not hold.
   Fig. \ref{SED}
   indicates that F1 reaches its maximum emission at a wavelength around 5 micron and then drops again
   and hence is not detected at 11.7 $\mu$m anymore. This suggests an object of lower luminosity
   (a few tens L$_{\odot}$, if really located at d = 5.7 kpc). 
   BN is 70 -- 100 times more luminous than F1 (1500 -- 2500 L$_{\odot}$, Dougados et al. \cite{Dougados},
   Gezari et al. \cite{Gezari1}), it has probably a far more massive reservoir of dust 
   around and is of course still strong at 10 and 20 $\mu$m.

   \subsection{Comparison with other HMCs}
    

   To date, the number of firmly established HMC sources is still quite small.  Apart from the 
   well--investigated Orion BN--KL region, only a 
   few HMCs have been studied by means of high--resolution infrared observations (cf.
   Stecklum et al. \cite{Stecklum1}, \cite{Stecklum2}, De Buizer et al. \cite{Buizer3}, \cite{Buizer4}, 
   Pascucci et al. \cite{Pascucci}).

   In this respect, it is worth comparing the infrared properties of our region with 
   a similar examination of the  hot molecular core in W3(H$_2$O). Both HMC regions seem 
   to be related to outflow activity. But in the latter case no NIR/MIR emission could be detected 
   in the direct vicinity of the HMC (see Stecklum et al. \cite{Stecklum2}). One explanation could be that the outflow 
   is mainly in the plane of the sky, so that we cannot benefit from an outflow-related clearing effect. \\
   Another possibility might be that the potential outflow material had
   not yet have time enough to sufficiently penetrate the very dense shell of the inner HMC 
   region as it is  instead apparently the case for our HMC component F.
   In that case, we would rate the G9.62+0.19 hot molecular core to be in a more evolved state than the 
   W3 one. 
   
   Support for this argumentation comes from recent 3-mm and 1.3-mm interferometric line observations of G9.62+0.19 
   with OVRO (Liu \cite{Liu}). This author discusses the use of several molecules as chemical clocks. 
   First, he uses H$_2$S to re--confirm the age differentiation among the components D, E, and F,
   since H$_2$S is eventually transformed into other species after it has been liberated from grain mantles
   into the warm molecular core environment. Among the three cores, component F shows the highest
   H$_2$S abundance and would thus be younger than E and D which both have already developed an ultracompact
   H{\sc ii} region. This finding is in accordance with our previous results 
   of the NH$_3$/CH$_3$CN abundance comparison (Hofner et al. \cite{Hofner2}). Furthermore, Liu 
   (\cite{Liu}) compares
   the CH$_3$OH and C$_2$H$_5$CN emission at 3 mm. While the observed methanol emission
   shows extended structure along the molecular ridge including D, E and F, ethyl cyanide is primarily
   detected toward core F. Since revised chemical HMC models usually produce significant amounts of
   N--bearing molecules only after O--bearing species have reached their peak abundances 
   (e.g., Rodgers \& Charnley \cite{Rodgers1}, \cite{Rodgers2}), the strong 
   C$_2$H$_5$CN emission toward component F suggests that it is in a more advanced stage of HMC
   evolution.
   
   This fits well to the scenario advocated by Kurtz et al. (\cite{Kurtz1})
   that HMCs can evolve first to hypercompact (Gaume et al. \cite{Gaume}, Tieftrunk et al. \cite{Tieftrunk}, 
   Kurtz \cite{Kurtz2}) and finally to ultracompact H{\small II} regions.
   The detection of weak emission at 1.3 and 3.6 cm, associated with a high emission measure of 
   $3 \times 10^8$ pc cm$^{-6}$, by Testi et al. (\cite{Testi2}) implies that the HMC component F
   is on its way to cultivate a hypercompact H{\small II} region. \\
    Furthermore, Cesaroni et al.
   (\cite{Cesaroni2}) analysed two other well--known hot molecular cores (G10.47+0.03 and G31.41+0.31) 
   and found indications that the HMC stage
   marks the phase of transformation from spherically symmetric to flattened and non--spherical
   structures. Also in this regard, G9.62+0.19--F proves to be a typical example of an HMC.

  \subsection{Why can't we see the D and E components in the infrared?}\label{why}

   Radio component F which is only a weak cm continuum source
   nevertheless shows infrared features in its immediate vicinity. Hence, at first glance it might be
   intriguing that the stronger radio sources D and E, which both are
   more massive and energetic than component F (Hofner et al. \cite{Hofner2}), do not show an infrared counterpart.
   Without a further analysis one might come to a similar conclusion as De Buizer et al. (\cite{Buizer}) who, for lack of
   an exact astrometry, just shifted their mid--infrared source to the position of the radio D component.  
   To enlighten these circumstances, we reconsider radio data from previous papers about G9.62+0.19 in order 
   to derive the expected Brackett $\gamma \,$ fluxes from as well as the extinction towards D and E. 
      \begin{table*}
      \caption[]{Physical parameters of the radio components D and E}
         \label{brg}
    
  \hspace*{1.5cm}       \begin{tabular}{lrrl}
            \hline
	    \hline
            \noalign{\smallskip}
       Physical parameter                          &  component D               &  component E             &  extinction law $\, ^{\rm b}$\\
            \noalign{\smallskip}
            \hline
            \noalign{\smallskip}
       $T_{\rm b}$  [K] $\,\, ^{\rm a}$            &   977                      &   210                    &\\
       $T_{\rm e}$  [K]                            &  8000                      &  8000                    &\\
       $N$(H) [cm$^{-2}$]                          &  $2.54  \times 10^{24}$    &  $1.84 \times 10^{24}$   &\\
       $S_{\rm Br \gamma}$  [W cm$^{-2}$]          &  $1.083 \times 10^{-19}$   &  $2.21 \times 10^{-20}$  &\\
       $A_{\rm 2.2 \mu m}$  [mag]                  &   125  		        &	 90		   & $R_{\mathrm V} = 3.1$, $CF$($2.2\mu$m)=0.108\\
       $A_{\rm 12 \mu m}\,$ [mag]                  &    32  		        &	 23		   & $R_{\mathrm V} = 3.1$, $CF$($\,12 \mu$m)=0.028\\
       $A_{\rm 2.2 \mu m}$  [mag]                  &   232  		        &	168		   & $R_{\mathrm V} = 5.0$, $CF$($2.2\mu$m)=0.125\\
       $A_{\rm 12 \mu m}\,$ [mag]                  &    60  		        &	 43		   & $R_{\mathrm V} = 5.0$, $CF$($\,12 \mu$m)=0.032\\

            \noalign{\smallskip}
            \hline
         \end{tabular}
     
\begin{list}{}{}
\item[$^{\mathrm{a}}$] {The brightness temperatures were derived taking the emission in the inner 
                        0$.\!\!''$55 of the components into account which corresponds to our
			K$_{\rm s}$ band resolution element (i.e. seeing).}
\item[$^{\mathrm{b}}$] {see Sect. \ref{extinction} for details}			
\end{list}
   \end{table*} 
   
  \subsubsection{The expected Br$\,\gamma$ fluxes of D and E}
  
   The method to compute the expected flux in IR recombination lines on the basis of cm continuum data
   has been explained extensively by Watson et al. (\cite{Watson}) and was also applied in some of our previous 
   investigations of other massive star--forming regions (e.g., Feldt et al. \cite{Feldt1}, \cite{Feldt2}).
   We refer to these papers for details and start with their formula for the emission measure ($EM$),
   which can basically be derived from the approximation of Altenhoff et al. (\cite{Altenhoff}) for the optical depth of 
   cm free--free emission:
   \begin{equation}\label{EM}
       EM = 4.72 \,\, a^{-1} \,\, T_{\mathrm{e}}^{1.35} \,\, \nu^{2.1} \,\, \mathrm{ln} \,\Big( \frac{T_{\mathrm{e}}}
            {T_{\mathrm{e}} - T_{\mathrm{b}}} \Big)\; \mathrm{cm}^{-5} \;.
   \end{equation}
   Herein, $T_{\rm e}$ and $T_{\rm b}$ are the electron temperature and the brightness temperature, respectively, 
   both in Kelvin. The parameter $\nu$ is the frequency (in Hz) at which $T_{\rm b}$ was
   derived. Finally, $a$($\nu,T$) is a correction factor of the order of 1, which is tabulated in Mezger \& Henderson 
   (\cite{Mezger})
   for various frequencies and electron temperatures.\footnote{Beckert et al. (\cite{Beckert}) demonstrate that this
   approximation (Altenhoff $+$ Mezger\&Henderson correction factor) is in very good agreement with
   non--approximative numerical solutions up to frequencies of $\sim 100$ GHz.} 

    By adopting the approach of Osterbrock (\cite{Osterbrock}) we can use the emission measure to 
   compute the expected Br $\gamma$ flux under the assumption that no extinction occurs:
   \begin{equation}\label{Brg}
       S_{\small \mathrm{Br} \gamma} = 0.9 \,\, {\mathrm h}  \nu_{\, \small \mathrm{Br} \gamma}  \,\, 
       \alpha^{\small \mathrm{eff}}_{\small \mathrm{Br} \gamma} \,\,  \frac{\Omega_{\small \mathrm{Br} \gamma}}
            {4 \, \pi} \, \, EM \; .
   \end{equation}
   The factor 0.9 reflects the assumption that a UCH{\sc ii} region consists of $\sim$10 \% of 
   singly ionised Helium, which contributes electrons $and$ ions to the cm free--free emission but 
   only electrons to the Br $\gamma$ emission. For 
   $\alpha^{\small \mathrm{eff}}_{\small \mathrm{Br} \gamma}$,
   which comprises the transition coefficients and the level populations via a temperature dependency,
   Hummer \& Storey (\cite{Hummer}) give:
   \begin{equation}\label{alpha}
      \alpha^{\small \mathrm{eff}}_{\small \mathrm{Br} \gamma} = 6.48 \times 10^{-11} \, T_{\mathrm e}^{\,-1.06} \; .
   \end{equation}
   For our computations we derived the brightness temperatures of components D and E from the 1.3 cm
   map of Testi et al. (\cite{Testi2}).\footnote{The 3.6 cm map of Testi et al. (\cite{Testi2}) 
   has a higher SNR, but Hofner et al. (\cite{Hofner2}) found that
   the behaviour of the components D and E already deviates from the optically thin regime at this 
   wavelength.} 
   The electron temperatures of both regions were
   estimated by Hofner et al. (\cite{Hofner2}), using a thermal bremsstrahlung model to fit the cm continuum 
   observations conducted up to that point. We list all the parameters in Table \ref{brg}.   

   The expected Br$\gamma \,$ flux is $1.083 \times 10^{-19} \,$ W cm$^{-2}$ for component D and
   $2.21 \times 10^{-20} \,$ W cm$^{-2}$ for component E for the theoretical limit of zero extinction.
   These values correspond to 5.75 mag and 7.48 mag in terms of K band magnitudes.
   If we use the ISAAC camera in combination with the Br$\gamma \,$ filter, we would detect such
   fluxes with a SNR of 20 within less than 1 s of exposure time.

  \subsubsection{The extinction towards D and E}\label{extinction}

   To evaluate these numbers one has to estimate the extinction towards these regions. Hofner et al. 
   (\cite{Hofner2}) derived H$_2$ column densities, based on 
   interferometric C$^{18}$O(1--0) observations, for the components D, E, and F. They report values of
   $N$(H$_2$) = 12.7 $\times 10^{23} \,$ cm$^{-2}$ for component D and $N$(H$_2$) = 
   9.2 $\times 10^{23} \,$ cm$^{-2}$ for component E. To transform these column densities into extinction
   values we start with the empirical relation that links the total hydrogen column density $N$(H) to the 
   optical colour excess $E_{\, \mathrm B-V} \,$ (Ryter \cite{Ryter}):
    \begin{equation}\label{ryter1}
      N(\mathrm H) = 6.83 \times 10^{21} \, E_{\, \mathrm B-V} \; \mathrm{cm}^{-2} \; \mathrm{mag}^{-1} \; .
    \end{equation}
   In order to substitute  $E_{\, \mathrm B-V}$, we use the ratio $R_{\mathrm V}$ of absolute to relative 
   extinction:
    \begin{equation}\label{ryter2}
      R_{\mathrm V} = A_{\mathrm V} \, / \, E_{\, \mathrm B-V} \; .
    \end{equation}
   In combining Eqs. (\ref{ryter1}) and (\ref{ryter2}) we get an expression for the visible extinction 
   $A_{\mathrm V}$, based on column densities. Finally, we can introduce a conversion factor $CF$($\lambda$) 
   that represents the conversion from the visible extinction ($A_{\rm V}$) to extinction values at
   other wavelengths ($A_{\lambda}$). We arrive at the following expression:
    \begin{equation}\label{ryter3}
      A_{\lambda} = \frac{R_{\mathrm V} \; N\mathrm{(H)}}{6.83 \times 10^{21} \, \mathrm{cm}^{-2} \, 
         \mathrm{mag}^{-1}} \; CF\mathrm{(\lambda)} \; .
    \end{equation}
   Mathis (\cite{Mathis}) gives two sets of conversion factors, corresponding to two possible
   extinction laws: normal ``diffuse interstellar dust'' (R$_{\rm V} = 3.1$) or ``outer--cloud dust'' 
   (R$_{\rm V} = 5.0$). For both possibilities, we compute extinction values at 2.2 $\mu$m and 12 $\mu$m
   by using $N$(H) $ = 2 \times N$(H$_2$). The results are listed in Table \ref{brg}.
   In any case, the extinction is $\ge$ 90 mag at 2.2~$\mu$m and still $\ge$ 23 mag at 12~$\mu$m.
   Such a large extinction would attenuate the fluxes down to an
   undetectable value. But one important restriction applies to this ``na$\ddot{\i}$ve'' interpretation of
   the extinction data. The C$^{18}$O line is optically thin; hence, it traces the column density
   of the whole molecular cloud and not just the column density up to the UCH{\sc ii} regions.
   Theoretically, the UCH{\sc ii} regions could even be located in front of the molecular clouds,
   which is obviously not the case.
   
  \subsubsection{Relative positions of UCH{\sc ii} regions and the molecular gas}
   
   One approach to clarify the situation is to compare the LSR velocities of radio recombination
   lines (RRLs) produced in the UCH{\sc ii} region with the LSR velocities of molecular
   lines seen in absorption against the radio continuum of the UCH{\sc ii} region. This idea has
   already been  pursued by Downes et al. (\cite{Downes}) and was recently revived by Araya et al. (\cite{Araya})
   who could show
   that in almost all of their UCH{\sc ii} region targets the ionised gas and the main absorbing
   molecular component have the same velocity within the uncertainties of the measurements. This demonstrates that
   a close association between the two classes of objects is very common. \\
   1.) 
   As far as we know, no radio recombination line studies exist for G9.62+0.19
   that would provide high spatial resolution. Therefore, we go back to the seminal work of Downes et al. 
   (\cite{Downes}) who
   conducted a combined survey of the H110$\alpha$ RRL and H$_2$CO (1$_{10}-1_{11}$) absorption line
   with the Effelsberg 100 m telescope at 6 cm. Their observations for G9.62+0.19 were centred on component B 
   and had a half--power beamwidth of $\sim 2.6$ arcmin. The reported LSR velocities are 
   ($3.0 \, \pm 5.0$) km s$^{-1}$ for the RRL and ($2.0 \, \pm 0.5$) km s$^{-1}$ for the H$_2$CO absorption
   line.\footnote{The RRL velocity is confirmed by Lockman (\cite{Lockman}) who gives $v_{\rm LSR}$ = 
   ($4.1 \, \pm 0.8$) km s$^{-1}$ for the H85$\alpha$, H87$\alpha$ and H88$\alpha$ lines towards 
   G9.62+0.19, based on observations with the NRAO 43 m telescope with  a FWHP of 3 arcmin.} 
   The velocities are almost equal which speaks for a close association also in the case of G9.62+0.19. \\
   2.) The fact that we actually see absorption implies that a noticeable fraction of the
   molecular gas is located in front of the H{\sc ii} regions. \\
   3.) The measured H$_2$CO line temperature for G9.62+0.19 is smaller than the continuum temperature 
   of the ionised sources (Downes et al. \cite{Downes}). An opposite finding would imply that most of the 
   H$_2$CO absorption occurs just by absorbing photons of the Cosmic Microwave Background and would 
   therefore not be related to the H{\sc ii} regions at all. \\
   4.) In our HCO$^+$ data of G9.62+0.19 (Hofner et al. \cite{Hofner4}), which we took with the IRAM 30 m telescope 
   and with the Plateau de Bure (PdB) interferometer, we clearly see a strong absorption dip in this molecular 
   line at the same LSR velocity as the RRLs and the H$_2$CO absorption occur. An interpretation, we 
   mentioned in the paper, is self--absorption due to a high optical depth. However, this feature can
   also be understood as absorption against the 89 GHz continuum of the (ultra)compact H{\sc ii} regions.
   Probably, both effects contribute to the observed line absorption.
   Note, that the PdB data provide sufficient spatial resolution to distinguish between the 
   components D and E and that component B is not dominating in these data. \\
   5.) Regarding the morphology of the components D and E we do not see features which could   
   explain these UCH{\sc ii} regions within the scope of blister/champagne flow models (Tenorio--Tagle \cite{Tenorio},
   Yorke \cite{Yorke}),
   which would automatically come along with the location of the UCH{\sc ii} regions very near the border
   of the embedding molecular cloud. \\
   
   These five points support the idea that both cm continuum components (D and E) are clearly embedded 
   within their  parental molecular cloud, although we cannot exactly estimate their relative positions. 
    We cite here recent investigations by Kim \& Koo (\cite{Kim}), based on molecular line
   observations towards  16 UCH{\sc II} regions. They indicate that the formation of
   massive stars, in general, takes not place near the surface of molecular clouds but in their interiors.
   In this regard, one has to take into account that already the positioning of the UCH{\sc ii} 
   regions D and E behind 20 \% of the total amount
   of molecular gas would be sufficient to explain our non--detections.
   On the basis of these considerations we conclude that the invisibility of the components D and E in 
   the near-- and mid--infrared is a reasonable finding.


\section{Conclusions}

    We have performed a comprehensive set of infrared observations for the G9.62+0.19 star--forming
    region. Narrow--band and broad--band observations from
    1  to 19 $\mu$m provided new information about this complex
    of young stellar objects. In the following we summarise our results. \\
    \parbox[t]{0.4cm}{\bf 1.)} The high resolution of our data revealed the detailed 
                          structure of the Hot Molecular Core region G9.62+0.19--F. The K band emission
			  found by Testi et al. (\cite{Testi1}) could be disentangled into three
			  distinct objects.
                          Our astrometry shows that among these three objects the dominating source F1 
                          is not coincident with the peak of the molecular line emission of the hot
			  molecular core. Instead, it is displaced by $\sim 1.7$ arcsecs, which translates
			  to roughly 10000 AU on a linear scale. 
                          We estimate that the second object F2 is probably a foreground star not intrinsically
                          related to the star--forming region. No reliable interpretation can be given
			  for the third very faint object F3 in the HMC region, since it is only
			  visible in the Ks and L' band.\\
    \parbox[t]{0.4cm}{\bf 2.)} Very near the  peak of the HMC molecular line emission and partly blended
                          by the foreground star we 
                          find faint diffuse emission at 2.16 $\mu$m. While the foreground star F2
			  quickly fades at wavelengths $\,>3 \, \mu$m,  this emission strongly increases in the
			  thermal infrared and finally reveals the presence of another object F4. 
                          We propose that object F4 is directly associated with the
                          HMC. This presents the first
			  detection of a hot molecular core at a wavelength as short as 3.8 $\mu$m.
			  We know that the HMC probably drives a molecular 
                          outflow roughly aligned with the line of sight (Hofner et al. \cite{Hofner4}).
			  Thus, the clearing effect of this outflow might allow us to look deeper into 
			  the infrared--emitting regions of the core environment. The results of our
			  K band polarimetry support the interpretation that the diffuse K band emission
			  arises due to scattering of light in the outflow cone.  Finally, our mid--infrared
			  data highlight object F4 as the most luminous object within the HMC region.
			  With the SED of F4 the range for several parameters of the HMC can be narrowed.\\
    \parbox[t]{0.4cm}{\bf 3.)} The combined results of the astrometry and radiative transfer computations
                          make it unlikely that the dominant object F1 is a direct trace of the
			  HMC. Nevertheless, we have demonstrated that on the basis of the NIR colours and the
			  K band spectrum the most likely explanation
			  is a dense dust shell around a deeply embedded object of lower luminosity. This might be a hint for
			  multiplicity of YSOs within the HMC region.  In a way, G9.62+0.19-F resembles the appearance
			  of the Orion Hot Core region, where likewise the near vicinity of the HMC is
			  populated by a variety of infrared sources. \\
    \parbox[t]{0.4cm}{\bf 4.)}   We evalute the fact that the UCH{\sc II}s D and E are not detected
                          in our infrared data although these objects show much stronger radio emission than the 
			  HMC region F. By reexamining available high--resolution radio molecular line data
			  we derive theoretical values for the extinction towards D and E and show that our
			  non--detections are in agreement with these predictions.  \\
    \parbox[t]{0.4cm}{\bf 5.)} Using our K--narrow--band results and the imaging polarimetry we reveal 
                          within the whole G9.62+0.19 star--forming complex
                          well--defined regions of enhanced Br$\gamma$ and H$_2$ emission as well as a sector 
			  where a large contribution comes from scattered light.  This demonstrates the complex
			  composition of such high--mass star--formation regions that would escape our views
			  when just using simple broad--band imaging.\\			  			  		  	  			  
    \parbox[t]{0.4cm}{\bf 6.)} Generally, we come to the following conclusions. 
                           The infrared appearance of complex star forming regions is not exclusively governed
			  by the intrinsic spectral energy distribution of the YSOs but often dominated by the distribution
			  and asymmetries of the extincting circumstellar material on several scales. The interaction
			  of outflows from YSOs with their surroundings can lead to more complicated configurations.
                          In particular, infrared emission associated with HMCs is detectable under certain 
			  favourable circumstances, as we demonstrated for G9.62+0.19--F. \\
			  While spatial low--resolution data,
			  e.g., from the previous generation of IR satellites, could act as a principle guide line
			  for investigating massive star formation, high--resolution observations
			  are an essential tool to disentangle the objects in high--mass star--forming regions
			  and to make any fundamental progress in that field. In this respect, an accurate
			  astrometry is needed since it can have a strong impact on the interpretation
			  of the data. \bigskip \\

\noindent
{ \bf Acknowledgements} \smallskip\\
       We thank L.~Testi for providing the SOFI K band spectrum, A.~Watson for a helpful advice, and 
      J.~De Buizer for some stimulating discussions during two conference occasions. 
      We especially thank  our referee, Dr. Riccardo Cesaroni, for a constructive
      report that focussed the paper at some critical points where the referee seemed to have more 
      confidence in our work than we ourselves. 
      The work of H.~L. and B.~S. was supported by the German
      \emph{Deut\-sche For\-schungs\-ge\-mein\-schaft, DFG\/} project
      numbers Ste 605/17--1 and 605/17--2. 
      P.~H. acknowledges the partial support from the Research Corporation grant Nr. CC4996, as well as from
      NSF grant AST-0098524.
      This research has made extensive use of the {\sl ALADIN} interactive sky atlas
      (Bonnarel et al. \cite{Bonnarel}) and the {\sl VIZIER} catalogues (Ochsenbein et al.  \cite{Ochsenbein}), 
      both provided by {\sl CDS}, Strasbourg, France. Furthermore, we used NASA's Astrophysics Data System 
      (ADS) for accessing the literature listed in the references. \\
      This research made use of data products from the Midcourse Space Experiment ({\sl MSX}) and of the 
      NASA/IPAC Infrared Science Archive, which is operated by the Jet Propulsion Laboratory, California 
      Institute of Technology, under contract with the National Aeronautics and Space Administration.
      \vspace*{3cm}


\appendix


\section{NIR Astrometry}\label{astrometry}

   Special care was taken to establish an accurate astrometry for the {\sl VLT} NIR data. 
   Two approaches were used: \\
   First, we extracted an I--band image from the {\sl ALADIN} interactive sky atlas 
   (Bonnarel et al. \cite{Bonnarel}) that contained the region of interest. The image 
   we used had been constructed by digital scanning of  a {\sl SERC}--I photo plate, 
   using the {\sl MAMA/CAI} scanning facility (Guibert \cite{MAMA})). Such an image has -- in 
   comparison to {\sl DSS1} and {\sl DSS2} images -- a finer pixel scale (0$.\!\!''67$/px). 
   Within this ($11.\!\!'5 \times 11.\!\!'5$) subimage we found three useful stars also 
   included in the {\sl ACT} reference catalogue (Urban et al. \cite{Urban}). We calibrated 
   the I--band image with help of these {\sl ACT} stars and used this image now as an astrometric
   reference to adjust the {\sl VLT}--K$_{\mathrm s}$ image. This could be done straightforward 
   since many stars in the central region of the I--band image were also present
   in the {\sl VLT} image. 
   
   The second way was more direct without an intermediate step. By applying the 
   {\sl ALADIN} tool we identified more than ten stars in our {\sl VLT} image also included
   in the {\sl USNO-A2.0} catalogue (Monet et al. \cite{Monet})\footnote{ We used the USNO-A2.0
   catalogue because it is more or less directly tied to the HIPPARCOS reference frame. 
   In the meantime, we also checked the astrometry for the G9.62+0.19 region according to the
   2MASS All Sky Survey (Cutri et al. \cite{Cutri}). An offset of around 0.3 arcseconds seems
   to exist between both catalogues for G9.62+0.19. We can and will not evaluate here which 
   catalogue is more accurate. We just mention that the 2MASS astrometry places our object
   F4 even more close ($0\,$\farcs15) to the peak of the HMC molecular line emission 
   (see Fig \ref{gallery})}. We then used these catalogued stellar 
   positions to directly calibrate the {\sl VLT} image. Nevertheless, one has to be careful
   when choosing these stars. Some objects which, due to the lower resolution,
   seem to be star--like on the optical photo plate images and are therefore included
   in star catalogues like {\sl USNO-A2.0}, turn out to be stars with associated nebulosity
   or clusters of stars when imaged with high resolution. Hence, we controled our selection
   of stars by examining public available digitised images of the region. Only positions
   which appeared to be clearly associated with one single object in both, the optical and 
   the NIR images, have been used for the calibration. 
   
   Both methods gave almost identical results. A comparison of several object positions
   resulted in deviations of $\le 0.\!\!''25$ between the two methods. 
   
   After applying these astrometrical calibrations the 
   ``red'' compact object F1 is still 1.7 arcsec away from the position of radio component F --
   a discrepancy which we find to be significant regarding the high resolution of
   the data. Furthermore, we should mention that the positional difference between our astrometry 
   and the one in Testi et al. (\cite{Testi1}) is obvious, but not large. On the other hand, there
   is a substantial astrometric disagreement between our newly derived positions and the ones 
   given in Walsh et al. (\cite{Walsh2}) for their K band image of the G9.62+0.19 region. 
   
   Persi et al. (\cite{Persi}) use for their investigations of G9.62+0.19 the 
   (preliminary) {\sl DENIS} point source 
   catalogue (Epchtein et al. \cite{Epchtein}) as astrometric reference. This
   choice is problematic since this catalogue itself uses the old 
   (pre--{\sl HIPPARCOS}) Guide Star Catalog GSC~1 (Lasker et al. \cite{Lasker})
   for calibration. As a test, we looked for stars in our VLT images which 
   are included in USNO2 as well as in {\sl DENIS}.
   Simple marking of the coordinates given in the catalogues reveals that all
   {\sl DENIS} positions seem to be systematically shifted $\approx$ 2.5 arcsec
   to the north--northwest. From that spot check we cannot assert a global
   problem for the astrometry of the {\sl DENIS} catalogue. But it is obvious
   that the particular stripe of the {\sl DENIS} release data which contains 
   the G9.62+0.19 region suffers from an unprecise astrometry and should not be
   used for calibrating high--resolution data.\\


\section{Correct position of the compact MIR source in G9.62+0.19--F}\label{astrometry2}

   De Buizer et al. (\cite{Buizer}) had first reported on this MIR feature. They used previous 
   23 GHz continuum data (Cesaroni et al. \cite{Cesaroni1}) to correlate their MIR data. These radio data did not yet 
   reveal the presence of the weak 1.3 cm radio emission arising from component F, but only showed 
   the peak belonging to component D (cf. the overlaid contours in Fig. \ref{SC10}).
   They then assumed that this radio continuum peak for D should be coincident with the MIR peak. 
   However, more recent investigations could already
   show that the compact mid--IR source does not coincide with radio component D (see Stecklum et al. 
   \cite{Stecklum1}, Persi et al. \cite{Persi}). Both groups placed the emission within the HMC region
   G9.62+0.19--F. The astrometry was mainly based on the comparison of centroids of the extended emission 
   (component B) from the science data and from the (coarser) 12.1 $\mu$m {\sl MSX} images showing the 
   same field. Still, the error
   of such a method is around 2$''$ and does not allow to decide to which of the objects in the 
   HMC region the MIR compact emission is related. Finally, De Buizer et al. (\cite{Buizer4}) presented new
   11.7 $\mu$m data that seem to show that the MIR emission is neither associated with component D nor with
   component F but is displaced from both by roughly 2 -- 3 arcseconds. Their new astrometry is based on a
   correlation between MIR features and a 3.5 cm ATCA interferometric radio map taken from Phillips et al. 
   (\cite{Phillips}).\\
   In an attempt to clarify the situation, we observed G9.62+0.19 at 8.7 $\mu$m (N1 filter), using TIMMI2. The smaller
   wavelength (compared to 11.7 $\mu$m) and the larger field of view (compared to SpectroCam-10) should 
   facilitate the search for compact reference objects in the field. The result is shown in Fig. \ref{TIMMI2},
   where we overlay the TIMMI2 data on our NIR data. In the north, we found a bright star, that is also
   visible in the K$_{\rm s}$, L$'$, and M band, suitable for attaching the 
   TIMMI2 astrometry to the VLT astrometry. The small rotation misalignment of the TIMMI2 chip with respect to the sky coordinate
   system could be estimated independently, because during the observating run we also observed other regions
   that showed several sources with 2MASS counterparts in the field of view (Grady et al. \cite{Grady}, 
   Linz et al. \cite{Linz}). The positional uncertainty is around 0.4 arcsecs in RA and DEC
   (i.e. 2 pixels in each direction). We cannot confirm the astrometric result of 
   De Buizer et al. (\cite{Buizer4}). We find, that the compact MIR emission is indeed located in the 
   direct vicinity of the HMC. Our astrometry speaks against a straight identification of the compact MIR 
   emission with object F1, but rather suggests that the peak of the MIR emission is more towards object F4.    
   In comparison to the 11.7 $\mu$m data, the MIR emission blob is not pointlike.
   We therefore deconvolved this part of the image with the standard star HD 169916 acting as PSF reference.
   The result (cf. in Fig. \ref{gallery}) is an elongated structure with a similar position angle 
   ($55^{\circ} \pm 6^{\circ}$) as the
   F1--F4 object pair ($51.5^{\circ} \pm 2^{\circ})$. This further strengthens our diagnosis that also at 
   8.7 $\mu$m we see the objects F4 and F1 that seem to merge to one object due to the lower spatial 
   resolution of the the TIMMI2 data. The relative strengths of the contributions of both objects is
   governed by the very different spectral energy distributions of F1 and F4 (see Fig. \ref{SED}). \\
   We should finally mention that one fraction of the astrometrical disagreement between De Buizer et al. 
   (\cite{Buizer4}) and our result disappears when agreeing about the exact radio reference position for the HMC! 
   We note that they use
   the J2000 position $18^{\rm h} 06^{\rm m} 14\fsecs 82$ (RA), $-20^{\circ} 31' 38\farcs 4$ (DEC) as HMC reference position (cf. their Table 1 and
   their Fig. 2). But this corresponds neither to the peak of the NH$_3$(5,5) emission nor to the 
   position of the faint 3.6 cm continuum of the HMC (see our Table \ref{pos} or consult the respective 
   references therein). The difference between the 3.6 cm peak position, which we use as HMC position
   throughout this paper and which is the most accurate position for this hot core to date, and the De Buizer values is
   ca. 0.85 arcsec in RA and ca. 1.0 arcsec in DEC, the total displacement is hence around 1.3 arcsec.
   Taking this into account the disagreement between De Buizer's and our estimation still  does not vanish
   completely, but has shrunk considerably.


\section{MIR image restoration}\label{Landweber}

   If one has, due to the necessary chopping and nodding in the MIR, the positive and negative
   beams in the final image the usual way to treat the data is to perform a kind of shift and add.
   But in principle, it is possible to deconvolve the image using a matrix multiplication procedure.
   We refer to Bertero et al. (\cite{Bertero}; hereafter BBR) for details of this so--called projected 
   Landweber method. The
   basic idea is that within the final multi--beam image several points at different sky positions
   contribute to the signal contained in one pixel of the detector array. Assuming that the 
   chopping and nodding throws are colinear along the vertical orientation of the detector chip
   one can describe the multi--beam image $\mbox{\boldmath$g$}$ as result of a multiplication of 
   the actually mapped larger rectangular field $\mbox{\boldmath$f$}$ with a simple imaging matrix 
   {\bf A}, populated just with natural numbers (see BBR): 
   \begin{equation}\label{equ1}
       g = \mathrm{A} \, f \; .
   \end{equation}
   The projected Landweber method deals with this inversion problem in order to restore the
   original field $\mbox{\boldmath$f$}$. The following iterative ansatz is used:
   \begin{equation}\label{equ2}
   f^{\mathrm{(k+1)}} = \mathrm{P}_+ \left[ f^{\mathrm{(k)}} + \tau\left(\mathrm{A^T} g - 
   \mathrm{A^T A}f^{\mathrm{(k)}}\right) \right] \; .
   \end{equation}
   Hereby P$_+$ is the projection onto the positive subset of iterative solutions, 
   {\bf A}$^{\mathrm{T}}$ means the transposed matrix, and $\tau$ is a relaxation parameter 
   which is related to the largest singular value of the matrix {\bf A}. As starting point
   $\mbox{\boldmath $f$}^{(0)} = $ {\bf 0} is chosen.\\
   Applying this method, some effects can influence the quality of the restoration. This contains
   the appearance of ghost images of the restored objects at distances being multiples of the chopping/
   nodding throws. Such artefacts are inherent in the method itself and can be understood 
   mathematically. The BBR article deals with techniques to reduce/remove several kinds of artifacts
   that may appear.\\
   Since we used for our observations a chop--nod pattern different from the one used in the 
   BBR approach (chopping perpendicular to nodding vs. chopping parallel to nodding),
   we had to adapt the original version of their program. Due to the loss of colinearity the problem is
   now really a two--dimensional one; thus, one imaging matrix alone is not sufficient. The new imaging
   equation reads now with two imaging matrices {\bf A} and {\bf B}:
   \begin{equation}\label{equ3}
       g = \mathrm{A} \, f \, \, \mathrm{B} \; .
   \end{equation}
   The iteration equation evolves to the new key formula:
   \begin{equation}\label{equ4}
   f^{\mathrm{(k+1)}} = \mathrm{P}_+ \left[ f^{\mathrm{(k)}} + \tau \left(\mathrm{A^T} g \, \mathrm{B^T}- 
   \mathrm{A^T A} \, f^{\mathrm{(k)}} \, \mathrm{B \, B^T} \right) \right] \; .
   \end{equation}
   In principle, there is good reason to apply such a kind of restoration method.
   First, one can easily disentangle the regions in the multi--beam image where positive and
   negative components overlap when imaging crowded fields. In this regard,
   our approach has already proven to be effective for the case of TIMMI2
   imaging of NGC2264 IRS1 (Schreyer et al. \cite{Schreyer}).
   Second, one gains a larger field of view, because the reconstructed image covers the areas
   of both, the on--beams and the off--beams. This is of course an advantage when one tries to
   map extended objects with the medium--sized (not to say small) MIR detector arrays available 
   nowadays. In addition, within a larger field it might also be easier to find other objects
   that could serve as astrometric reference points. \\

\clearpage

\clearpage

\clearpage

\clearpage

\clearpage

\clearpage

\begin{figure*}
   \centering
   \includegraphics[width=14.75cm]{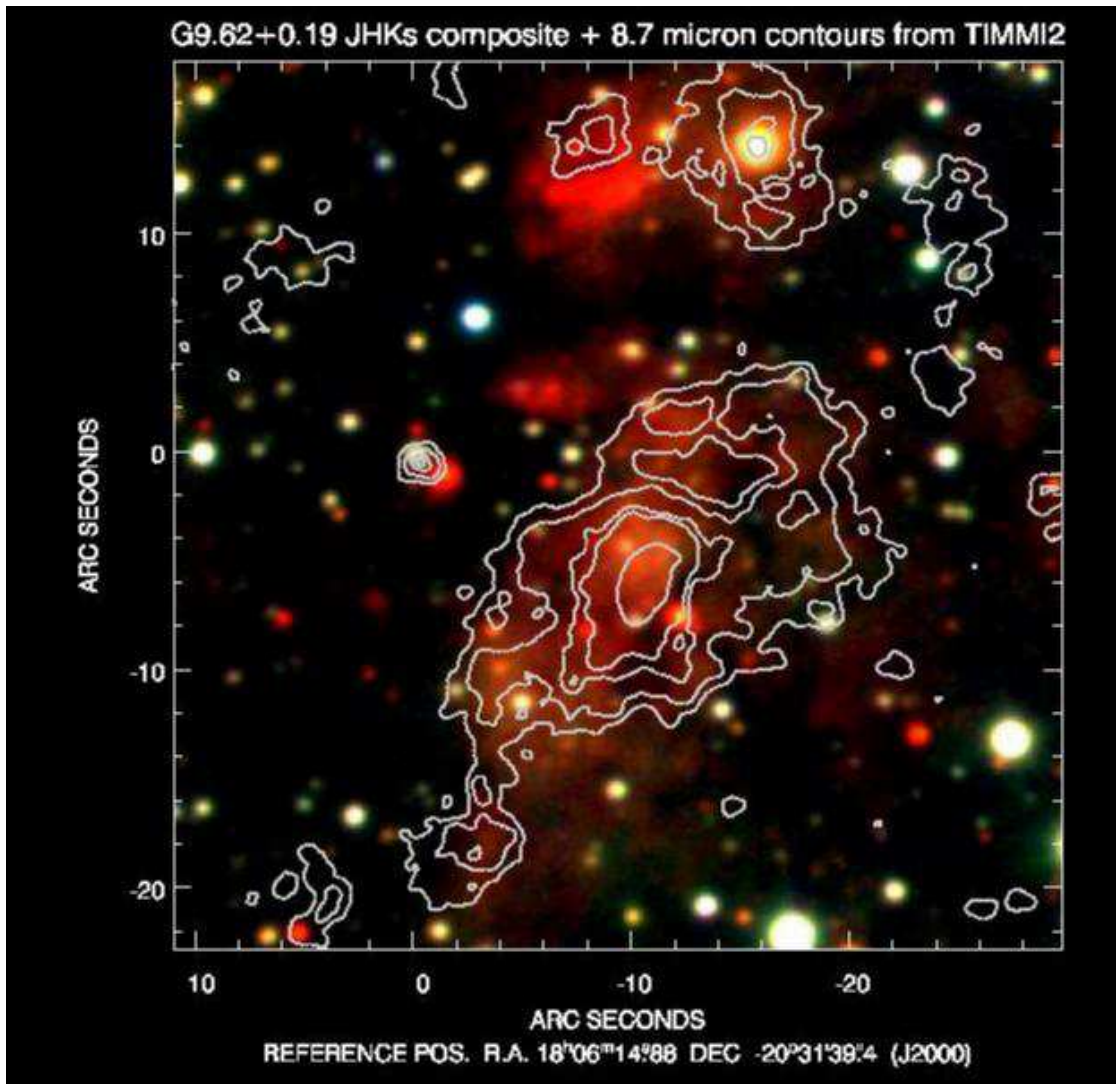}
      \caption{
	       The JHK$_{\rm s}$ colour composite, overlaid with the
	       TIMMI2 N1 band data as contours. The bright source at the offset coordinates ($-$16, 14)
	       served as main astrometric reference for the TIMMI2 data.
              }
         \label{TIMMI2}
   \end{figure*}

\clearpage

   \begin{figure*}
   \centering
   \includegraphics[width=\textwidth]{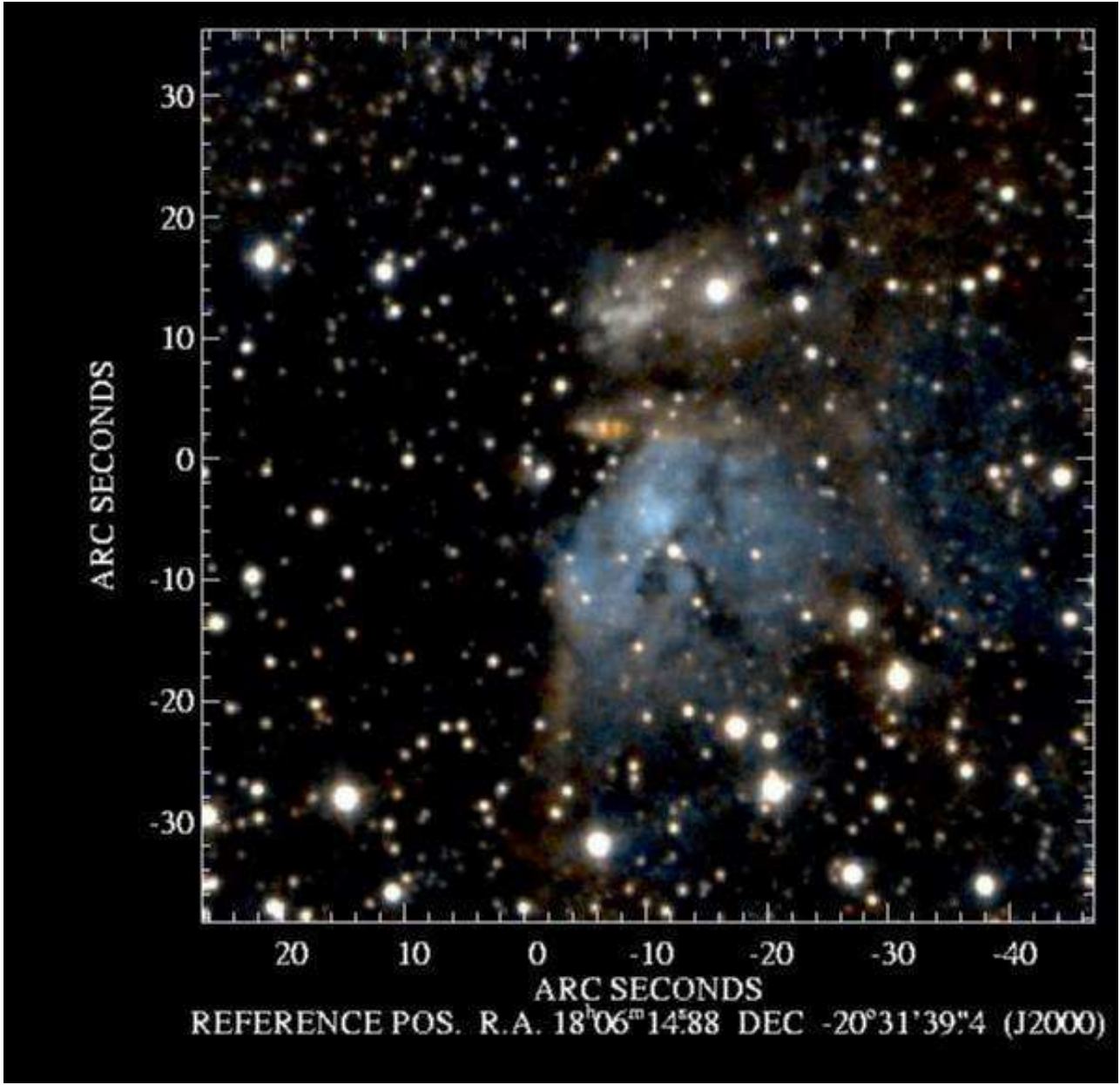}
      \caption{This image shows G9.62+0.19 in the K narrow--band filters: Br$\,\gamma$ (blue),
               H$_2$ (red), (Br $\gamma$ + H$_2$)/2 (green). The Br$\,\gamma$ emission arises mainly from
	       the compact H{\sc ii} region G9.62+0.19 B. Note that the H$_2$ emission blob roughly in the
	       image centre has its
	       own substructure. The diffuse emission to the north of it (white) turns out to be
	       mainly scattered light (see Fig. \ref{VLT4}).
               }
         \label{Brg_H2}
   \end{figure*}

\clearpage


   \begin{figure*}
   \centering
     \includegraphics[width=\textwidth]{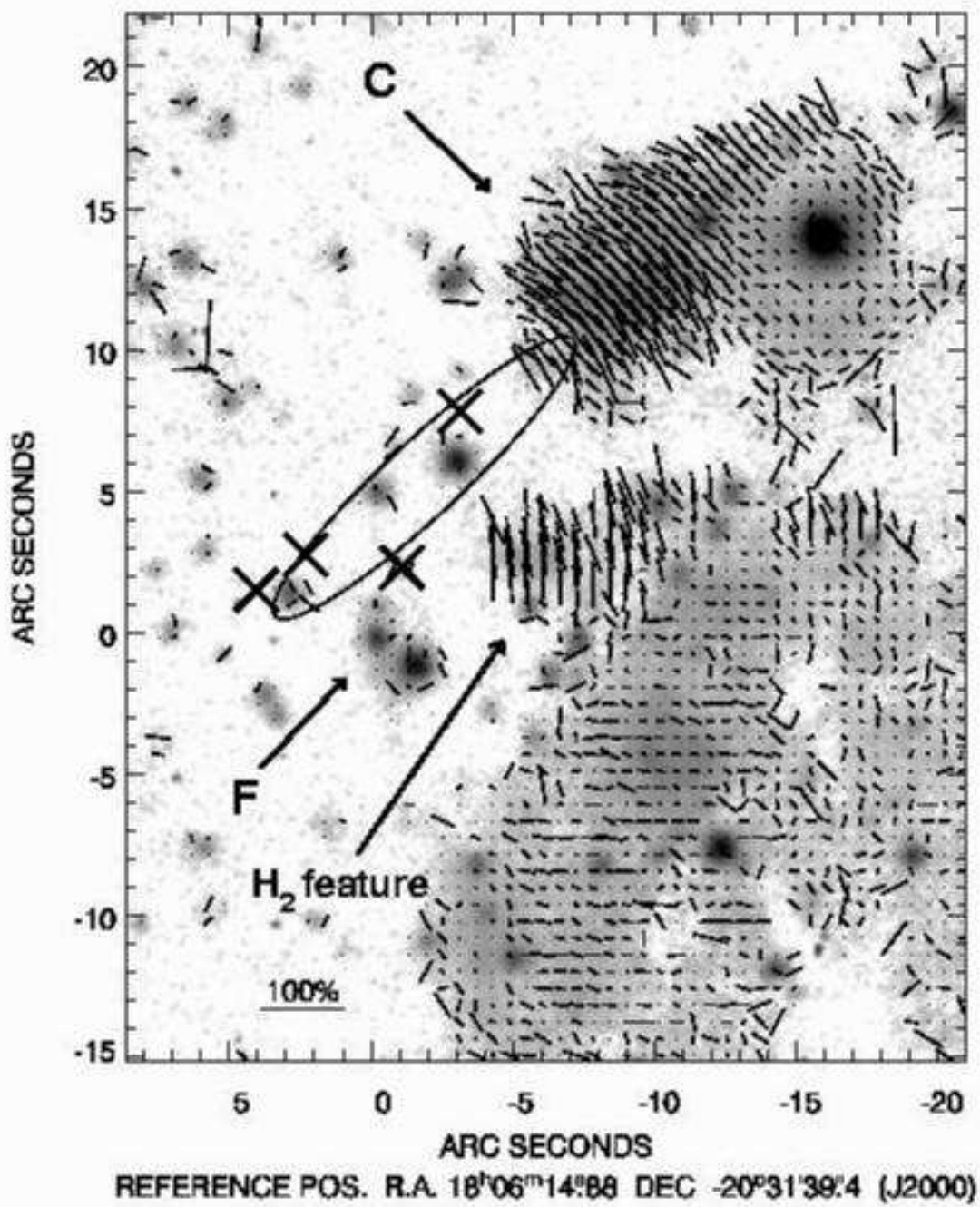}
      \caption{Polarisation map overlaid over an image of Stokes--$I$ (i.e. the intensity) of G9.62+0.19,
      both measured at 2.09 $\mu$m. The data were binned to a pixel size of $0 \farcs 6 \times 0 \farcs 6$ before
      deriving the polarisation. The ellipse is an indicator for the location of the illuminator whose scattered
      light causes the high degree of linear polarisation in some parts of the image  (see section
      \ref{pol data} for details).  The thick crosses mark the positions of the hypercompact
      H{\sc ii} regions E, G, H, and I (cf. Fig. \ref{VLT2}).}
         \label{VLT4}
   \end{figure*}

\clearpage

   \begin{figure}
   \centering
   \includegraphics[width=12.75cm]{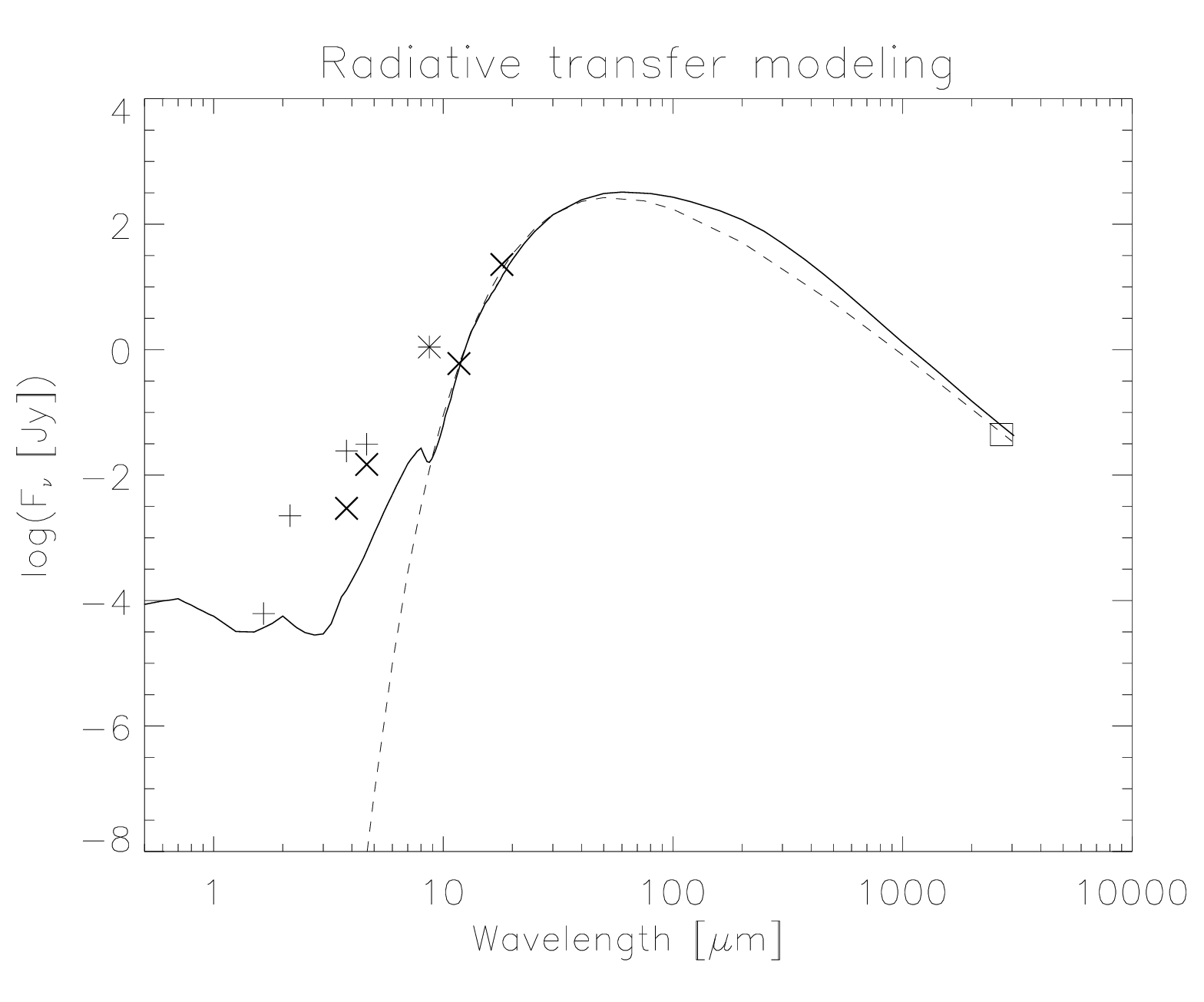}
      \caption{Example for the appearance of the SED of a hot molecular core, calculated with a radiative
               transfer model  (solid line). For this graph we used
               a central heating source with T$_{\mathrm{eff}} = $ 22500~K and a total 
               luminosity of 1.88 $\times 10^4$ L$_{\odot}$. The total mass contained
               in the model space was 95 M$_{\odot}$. 
	       For this particular model we assumed a constant density distribution and an outer radius
	       of 2600 AU. 
	       No accretion luminosity was added. The drawn--in
               symbols indicate the fluxes we measure for the objects F1 (plus--signs) and F4
	       (crosses), respectively. The asterisk at 8.7 $\mu$m denotes the integrated flux
	       from an area covering both F1 and F4. The 2.7 mm flux from Hofner et 
	       al. (\cite{Hofner2}) is marked by a small square.  As comparison, a 90 K 
	       modified blackbody is plotted as a dashed line. However, the significance of such a
	       temperature is put into perspective in Sect. \ref{RT}.
               }
         \label{SED}
   \end{figure} 
   
\clearpage

   \begin{figure}
   \centering
   \includegraphics[width=8.75cm]{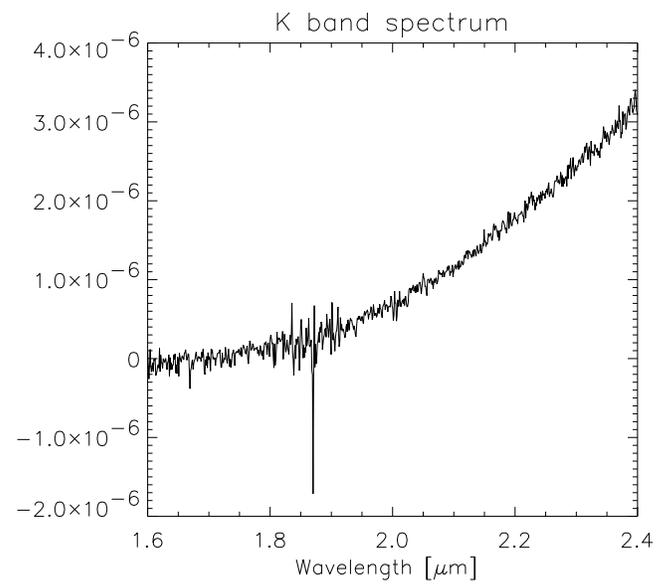}
      \caption{The K band medium--resolution spectrum of object F1 (in arbitrary units), taken with
      SOFI at the ESO NTT. 
               }
         \label{spectrum}
   \end{figure}

\end{document}